\documentclass[a4paper,11pt]{article}
\pdfoutput=1
\usepackage[T1]{fontenc}

\usepackage{jheppub} 
\usepackage{amsthm,amsmath,latexsym,amssymb,amsfonts,amscd}
\usepackage{graphics,lscape,fancyhdr,array,stmaryrd,euscript}
\usepackage{empheq}
\usepackage{tikz-cd}
\numberwithin{equation}{section}
\usepackage{hyperref}
\usepackage{ifthen,slashed}
\usepackage{bbm}
\usepackage{bbold}
\usepackage{float}
\usepackage{./diagrams/diagramsA}
\allowdisplaybreaks
\definecolor{linkblue}{rgb}{0.1,0.3,.7}
\definecolor{forestgreen(web)}{rgb}{0.13, 0.55, 0.13}
\definecolor{lava}{rgb}{0.81, 0.06, 0.13}
\hypersetup{
breaklinks,
    colorlinks,
    citecolor=forestgreen(web),
    filecolor=linkblue,
    linkcolor=lava,
    urlcolor=linkblue
}

\newcommand{\pl}{\partial}
\newcommand{\besubeqs}{\begin{subequations}}%
\newcommand{\esubeqs}{\end{subequations}}

\title{
\mbox{}\vspace{2cm} \\
{\LARGE \bfseries Quantum $\boldsymbol{\phi^4}$ theory in AdS${}_{\boldsymbol{4}}$ and its CFT dual
} }
\abstract{We compute the two- and four-point holographic correlation functions up to the second order in the coupling constant for a scalar $\phi^4$
theory in four-dimensional Euclidean anti--de Sitter space. 
Analytic expressions for the anomalous dimensions of the leading twist operators are found at one loop, both for Neumann and Dirichlet boundary conditions.\\


\noindent \textsc{Doi}: \href{https://doi.org/10.1007/JHEP02(2019)099}{10.1007/JHEP02(2019)099}}
\author[a]{Igor Bertan,}\emailAdd{igor.bertan@physik.lmu.de}
\author[a]{Ivo Sachs}\emailAdd{ivo.sachs@physik.lmu.de}
\author[b,c]{and Evgeny Skvortsov}\emailAdd{evgeny.skvortsov@aei.mpg.de}

\affiliation[a]{Arnold Sommerfeld Center for Theoretical Physics,\\ Ludwig Maximilian University of Munich,\\
Theresienstr. 37, D-80333 M\"unchen, Germany}
\affiliation[b]{Albert Einstein Institute,\\
Am M\"uhlenberg 1, D-14476, Potsdam-Golm, Germany}
\affiliation[c]{Lebedev Institute of Physics,\\
Leninsky ave. 53, 119991 Moscow, Russia}

\begin{document}
\pagestyle{empty}
\maketitle
\makeatletter
\renewcommand\section{\@startsection{section}{1}{\z@}%
                                   {-2.5ex \@plus -1.3ex \@minus -.7ex}%
                                   {2.0ex \@plus.4ex \@minus .4ex}%
                                   {\normalfont\large\bfseries}}
                                   \makeatother
\section{Introduction and summary}
\setcounter{page}{1}
When exploring quantum field theory (QFT) in curved geometries, de Sitter (dS) and anti--de Sitter (AdS) are the obvious space-times to consider since they are maximally symmetric, having the same number of isometries as Minkowski space-time. While dS is perhaps better motivated in cosmology as an approximation to the early \cite{Mukhanov:1981xt} and late time accelerated  \cite{Polyakov:2012uc} Universe, AdS has received much attention in mathematical physics due to the relation to conformal field theory (CFT) on its conformal boundary, as exemplified by the AdS/CFT correspondence \cite{Maldacena:1997re, Gubser:1998bc,Witten:1998qj}. On the other hand, progress in QFT on curved space-times has been hampered, in particular, by the absence of a momentum representation for which the Feynman amplitudes can be represented as elementary integrals. 
The absence of such a representation is particularly limiting at loop level. Indeed, while the short-distance properties of quantum fields in curved space-times can be analyzed systematically, little is known about the influence of curvature at distances of the order of the curvature scale. Consequently, to date, explicit tests of the AdS/CFT correspondence have, to a large extent, been limited to classical fields on AdS; that is, CFTs in the leading order in the $1/N$ expansion with a handful of examples and new techniques having just begun to appear to tackle the loop corrections, see, for instance, refs. \cite{Penedones:2010ue,Fitzpatrick:2011hu,Liu:2018jhs,Giombi:2017hpr,Aharony:2016dwx,Aprile:2017bgs,Alday:2017xua,Yuan:2018qva,Alday:2018pdi}.

Moreover, since the coupling of the CFT stress tensor to the graviton is present almost universally, one is confronted, when considering the bulk theory beyond the classical level, with the quantization of gravity together with its perturbative pathologies in the ultraviolet. Of course, when embedded in string theory, these singularities should be resolved, but world sheet calculations of string theory in AdS are mostly beyond reach at present \cite{Berkovits:2000fe}.

Another class of conjectured dualities, that was supposed to be simpler than the usual string-like dualities, is between CFTs with matter in vector representations of gauge groups and theories with massless higher spin fields in the bulk \cite{Klebanov:2002ja,Sezgin:2002rt}. The simplest example of such a duality is given by a free $O(N)$-vector model, i.e., a bunch of free scalar fields $\varphi^i(x)$, $i=1,...,N$ with the $O(N)$-singlet constraint imposed. This duality involves an infinite number of massless higher spin fields in the bulk that are dual to higher spin conserved tensors $J_s\sim \varphi \pl^s\varphi$ in the free scalar CFT. Other options include the critical vector model, the free fermion CFT, the Gross--Neveu model \cite{Sezgin:2003pt,Leigh:2003gk} and, more generally, Chern--Simons matter theories \cite{Giombi:2011kc}. The problem here is twofold. Firstly, higher spin theories reveal some pathological nonlocalities \cite{Bekaert:2015tva,Sleight:2017pcz,Ponomarev:2017qab} that prevent them from having a bulk definition that is independent of their CFT duals (but can be defined as anti--holographic duals of the corresponding CFTs). Secondly, for massless higher spin fields, ultraviolet pathologies of gravity are amplified with increasing spin, see ref. \cite{Ponomarev:2016jqk}. At present, it is unclear how they can be resolved, except for the conformal \cite{Fradkin:1985am,Tseytlin:2002gz,Segal:2002gd} and chiral \cite{Ponomarev:2016lrm} higher spin theories where the nonlocalities are absent and quantum corrections can be shown to vanish \cite{Beccaria:2016syk,Skvortsov:2018jea}.

One possible way to get around these problems, in the case of AdS, is to use the conformal bootstrap, in particular, crossing symmetry, to determine the coefficients in the operator product expansion (OPE) and the anomalous dimensions of the corresponding operators in CFT, to make predictions for loop-corrected boundary-to-boundary correlation functions of the dual bulk theory in AdS \cite{Heemskerk:2009pn,Aharony:2016dwx,Alday:2017xua,Aprile:2017bgs}. Here, the input is the first-order anomalous dimensions for the ``double-trace operators" inferred from the tree-level bulk amplitudes which, using crossing symmetry, lead to an equation for the second-order anomalous dimensions of the latter. This has led to explicit results for a class of dual bulk theories in
AdS${}_3$ and AdS${}_5 $ \cite{Aharony:2016dwx,Alday:2017xua,Aprile:2017bgs}. However, since there are no closed expressions for the conformal blocks in three  dimensions, this approach does not easily generalize to AdS${}_4$, which is the focus of the present work.

In the present paper we propose to compute loop-corrected correlation functions on the Poincar\'e patch of AdS without using any particular properties of a CFT dual on the conformal boundary \cite{Bertan:2018khc}.
More precisely, we consider the bulk theory in a loop expansion in position space. There is a convenient representation for loop diagrams in AdS in terms of Mellin amplitudes \cite{Penedones:2010ue}. A related approach, followed in ref. \cite{Fitzpatrick:2011hu}, is to reduce loop diagrams in global coordinates in AdS to a sum over tree-level diagrams using a discrete Mellin space K\"all\'en-Lehmann representation with weight function inferred from the OPE in the dual CFT.\footnote{We are not aware of an analogous construction on the Poincar\'e patch.} Alternatively, one may exploit the fact that the data defining the Mellin representation of the loop diagram is already contained in the tree-level data \cite{Aharony:2016dwx}.\footnote{We would like to thank E. Perlmutter for pointing this out to us.}

Here we will not follow this path. Instead, we simply evaluate the loop diagrams in an adapted representation in terms of Schwinger parameters with refinements due originally to Symanzik \cite{Symanzik:1972wj}. In order to avoid pathologies associated to spin-$2$ and above, we will consider a simple interacting scalar bulk theory. Concretely, we consider an interacting bulk scalar field with action
\begin{align*}
    S&=\int_{\mathrm{AdS}_4} \sqrt{g} \left( \frac12 (\partial\phi)^2 +\frac{m^2}{2} \phi^2+\frac{\lambda}{4!} \phi^4\right)
\end{align*}
on the Poincar{\'e} patch of Euclidean AdS${}_4$. This theory is perturbatively renormalizable and thus we will not have to deal with any of the pathologies mentioned above. In particular, we do not  quantize the bulk metric but instead treat it as a background. From the point of view of QFT on curved space-time this is a natural truncation.  
On the other hand, this may not seem so natural to a reader familiar with the AdS/CFT literature where the graviton appears naturally as the bulk field dual to the stress tensor of the CFT. However, if we do not insist on locality on the CFT side, there are many CFTs that do not possess a local stress tensor. Among them, the critical point of Ising-like models with long distance interactions, see, for example, ref. \cite{Behan:2017dwr}. Such CFTs should also admit an AdS dual description where gravity is frozen to a classical background.

The idea of truncating the bulk theory to an interacting scalar field is not new even in the context of the  AdS/CFT correspondence. In particular, this model was considered in ref. \cite{Heemskerk:2009pn} as a bulk dual to a CFT with just one low dimensional single-trace operator. The simplest such CFT is the generalized free field \cite{Greenberg:1961mr}, which is characterized by the property that the correlation functions of operators factorize similarly to those of the fundamental fields in the Gau\ss ian model. The corresponding bulk dual is just that of a free scalar field $\phi$ in AdS \cite{Witten:1998qj}. If we denote by $\mathcal{O}_\Delta$ the generalized free field of conformal dimension $\Delta$, then, using the standard AdS/CFT dictionary, its two-point function is given by 
\begin{equation}
 \langle \mathcal{O}_\Delta(x_1) \mathcal{O}_\Delta(x_2)\rangle_{\mathrm{CFT}}= \langle \bar{\phi}(x_1) \bar{\phi}(x_2) \rangle_{\mathrm{AdS}} = \frac{N_\phi}{r_{12}^{2\Delta}},
\end{equation}
where $\phi$ is the scalar field dual to $\mathcal{O}_\Delta$, $\bar\phi$ is its restriction to the boundary of AdS, $r_{12} \equiv |x_1 - x_2|$ and  $N_\phi$ is a normalization constant. Similarly the four-point function of the generalized free field will be given by 
\begin{equation}
 \langle \mathcal{O}_\Delta(x_1) \mathcal{O}_\Delta(x_3)\mathcal{O}_\Delta(x_4)\mathcal{O}_\Delta(x_2)\rangle_{\mathrm{CFT}}= \frac{N_\phi}{r_{12}^{2\Delta}}\frac{N_\phi}{r_{34}^{2\Delta}} \quad+\quad\text{permutations}.
\end{equation}
Crossing symmetry of the left hand side will be automatically satisfied and it then has an expansion in conformal blocks of the double-trace operators $\mathcal{O}_\Delta \square^n \pl^l \mathcal{O}_\Delta$ of the corresponding CFT.

Next, we consider a deformation of the generalized free field that does not preserve the factorization property. The simplest such renormalizable deformation is the $\phi^4$ theory on hyperbolic space. 
This deformation should correspond to an interacting CFT with a scalar operator $\mathcal{O}_\Delta$, $m^2\propto \Delta(\Delta-d)$, but without a local stress tensor.  
It is clear that any interaction term in an action for the bulk theory will give a crossing-symmetric contribution to the correlation functions on the CFT side by construction. At present, we will take the conformally coupled scalar field in AdS${}_4$. There are two possible choices of boundary conditions for $\phi$: $\Delta=2$ and $\Delta=1$, both being within the unitarity window \cite{Klebanov:1999tb}.   
Due to the extremality of the $\Delta=1$ case we expect some subtleties at the quantum level. For the same reason we do not include the $\phi^3$ interaction. We will discuss this in more detail throughout the text.

The first prediction for the CFT that is computable at tree-level in AdS is the anomalous dimensions and OPE coefficients of double-trace operators appearing in the OPE of $\mathcal{O}_\Delta$ with itself. This can be extracted from the exchanges and quartic contact interactions, see, for instance, refs. \cite{Hoffmann:2000mx,Arutyunov:2000ku,Heemskerk:2009pn,Giombi:2017cqn}. We then compute the first quantum corrections to the two- and four-point functions, which includes one- and two-loop diagrams in AdS${}_4$. At a conceptual level, an important implication of this is that the actual loop calculation in the bulk theory is consistent with the duality \cite{Bertan:2018khc}. In addition,  
this allows us to extract further CFT data. In particular, we can extract higher order corrections to the anomalous dimension of double-trace operators, as well as their OPE coefficients at next-to-leading order in both the deformation parameter $\lambda$ and the dimension of the double-trace operators. One result, already noted in ref. \cite{Heemskerk:2009pn}, is that, while the conformal block expansion of the four-point function of the generalized free field involves primary double-trace operators\footnote{In the present context, based on the identification obtained in the original AdS/CFT conjecture, we denote by double-trace operators all operators which are not dual to a bulk field.} of all even spin and even dimensions, only the OPE coefficients and dimensions of such operators with spin $0$ are affected by the interaction at tree-level.\footnote{This may come as a surprise since the higher spin primaries do not correspond to conserved currents as they do not saturate the unitarity bound.} At loop level, however, their dimensions are corrected (see also ref. \cite{Aharony:2016dwx} for AdS${}_3$ and AdS${}_5$). 

One of the main results of this paper is the anomalous dimensions $\Delta_{0,l}$ of the operators of the leading Regge trajectory, i.e., having the form $\mathcal{O}_\Delta \pl^l \mathcal{O}_\Delta$. These are the lowest-twist double-trace primaries appearing in the OPE of $\mathcal{O}_\Delta$ with itself. For $\Delta=2$ we find
\begin{align}
     \Delta_{0,l}=4+l+ \gamma\delta_{l,0} + \gamma^2
    \begin{cases}
        \frac53 & \mathrm{for~} l=0,\\
        -\frac{6}{ (l+3) (l+2)(l+1) l} & \mathrm{for~} l>0,
    \end{cases}
\end{align}
where $\gamma=-\lambda_R/16 \pi ^2$ and $\lambda_R$ is the renormalized coupling. For $ \Delta=1$ we have, in turn,
\begin{align}
   \Delta_{0,l}=2+l+2 \gamma\delta_{l,0} +\gamma^2 \frac{-4 }{2 l+1}\psi^{(1)}(l+1)+\gamma^2\begin{cases}
        -4 & \mathrm{for~} l=0,\\
        -\frac{2}{l(l+1) } & \mathrm{for~}l>0,
    \end{cases}
\end{align}
where $\psi^{(1)}(l+1)$ is the trigamma function. Anomalous dimensions for higher twist operators are also computed but they do not seem to have such a simple $l$-dependence. 

Our results have a direct bearing on higher spin theories in AdS${}_4$ as well.  Indeed, all these theories contain a scalar field corresponding to $\Delta=1,2$ and need to have a vanishing $\phi^3$ bulk coupling.\footnote{In higher spin theories one should add boundary terms in order to obtain the correct $\langle \mathcal{O}\mathcal{O}\mathcal{O}\rangle$ correlator \cite{Freedman:2016yue}. We will not consider these in the present paper. The interplay between bulk and boundary terms can lead to interesting effects for loops. } The quartic vertices begin with $\phi^4$ and contain infinitely many $(\phi\overleftrightarrow{\nabla}^k\phi)\square^n(\phi\overleftrightarrow{\nabla}^k\phi)$ vertices \cite{Bekaert:2015tva}. Therefore, our results present a meaningful contribution of the $\phi^4$ interaction to the anomalous dimensions and OPE coefficients of higher spin theories and show how to deal with ultraviolet and infrared divergences in the bulk. 

An important question that requires clarification at loop level in AdS concerns the dependence on the renormalization scheme. In this paper we use an ultraviolet cut-off regularization, which manifestly preserves covariance, followed by a nonminimal subtraction. For $\Delta=1$, there are additional infrared divergences. A convenient covariant regularization of the latter is provided by continuation in $\Delta$. 
Another issue related to this is the lack of a simple quantum AdS experiment that determines the renormalization conditions in terms of measurable quantities such as the mass of particles, for instance. In the present context we replace the latter by the dimensions of the operators of the dual CFT which seems to be an appropriate replacement in the context of AdS/CFT \cite{Bertan:2018khc}. 

In section \ref{prel} we specify our conventions and review the construction of the scalar propagator on AdS. In addition, we list the various bulk correlation functions that will be calculated later on.  

In section \ref{2ptfnchapter} we derive the one- and two-loop corrections to the bulk-to-bulk two-point function in position space using a Schwinger parameterization. Here we also specify the ultraviolet regularization employed in this paper. For $\Delta=1$, we will encounter in addition infrared divergencies  whose regularization is also discussed there. 

Section \ref{4ptfnchapter} contains the computation of the four-point function. The tree-level contribution is well known (e.g., refs. \cite{Arutyunov:2000ku,Heemskerk:2009pn}), so that we will just recall the result. The one-loop contribution requires an ingenious use of Schwinger parameters. Eventually, the ultraviolet divergences can be absorbed in the renormalized $\phi^4$ coupling as expected. It turns out that the $\Delta=1$ calculation differs from that for $\Delta=2$ by an extra contribution, which is computationally tedious but manageable as a short-distance expansion on the boundary. 

In section \ref{chapope} we then compare the short-distance expansion of the bulk four-point function with the conformal block expansion in conformal field theory. At zeroth order in the bulk coupling it is possible to read off the spectrum of double-trace operators. At order $\lambda$ one determines the anomalous dimensions which vanish for all but the spin-$0$ double-trace operators at that order. At order $\lambda^2$ things become more interesting. Still, for the leading Regge trajectory, we are able to derive a closed formula for the anomalous dimensions. 

Section \ref{conc} contains the conclusions. An extensive list of anomalous dimensions and OPE coefficients for various spins and twists are referred to the appendix.

\section{Preliminaries}\label{prel}
We briefly review the kinematical ingredients that are used throughout the paper. The $(d+1)$-dimensional Euclidean anti--de Sitter space,  $\mathbb{H}_{d+1}$, can be embedded into a $(d+2)$-dimensional flat Minkowski ambient space $\mathbb{M}_{d+2}$. The space $\mathbb{H}_{d+1}$  is then one of the sheets of the two-sheeted hyperboloid
\begin{equation}
\label{quadric}
X^2: = \eta_{AB} X^A X^B = (X^0)^2+...+(X^{d})^2-(X^{d+1})^2 = -\frac{1}{a^2},
\end{equation}
where $A=0,...,d+1$. Given two points $X$ and $Y$ on the hyperboloid, there is a simple relation between the geodesic distance $\rho$ and the scalar product of $X$ with $Y$
\begin{equation}
\label{geodist}
\cosh a \rho = - a^2 X \cdot Y.
\end{equation} 
A useful parameterization which, however, covers only half of the space, is given by the Poincar\'e coordinates
\begin{align}
\begin{split}
X^0 =& \frac{1}{\sqrt{2} a z} (1 - \frac{x^{i2}}{2} -\frac{z^2}{2}),\\
X^i =& \frac{x^i}{a z},\\
X^4 =& \frac{1}{\sqrt{2}a z} (1 + \frac{x^{i^2}}{2} +\frac{z^2}{2}),
\end{split}
\end{align}
where $z>0$ and $i=1,...,d$. In these coordinates the metric makes the  conformal flatness explicit
\begin{equation}
\label{poincmetricc}
\mathrm{d}s^2 =\frac{1}{a^2 z^2} (\mathrm{d}z^2 + \mathrm{d}{x^{i}}^2).
\end{equation}
We then introduce a dimensionless $O(d+1,1)$-invariant quantity, related to the geodesic distance (\ref{geodist}), $K:= -\frac{1}{a^2 X \cdot Y}$, which, when expressed in Poincar\'e coordinates, reads
\begin{equation}
\label{uinpoinc}
K = \frac{2zw}{(x^i - y^i)^2 + z^2 + w^2},
\end{equation}
where $x^\mu\equiv x =(z,x^i)$ and $y^\mu \equiv y = \left(w, y^i \right)$. The points at $z=0$ are said to be ``at the boundary" of anti--de Sitter space, and for further reference we note that in the limit where $z$ approaches $0$ we have
\begin{equation}
\label{boundbeh}
 K \sim z  \bar{K},
 \end{equation} 
 where
 \begin{equation}
 \label{witpro}
 \bar{K} = \frac{2w}{(x^i - y^i)^2 + w^2}
 \end{equation}
is the usual bulk-to-boundary propagator \cite{Witten:1998qj}. 

The flat space limit is obtained by letting  $a\rightarrow0$. By introducing spherical coordinates with the radial coordinate defined by $r = \sqrt{X^2 + X_{d+1}^2}$, $K$ can be shown to have the expansion
\begin{equation}
\label{flatspacelimitofK}
K  \equiv \frac{1}{\sqrt{a^2 r^2 +1}}= 1-  \frac{1}{2}a^2 r^2+ \mathcal{O}(a^4).
\end{equation}

In order to evaluate the Feynman bulk-to-bulk  diagrams we need the bulk-to-bulk propagator $\Lambda(x,y;m)$ for the scalar field.  By definition, $\Lambda(x,y;m)$ satisfies
\begin{equation}
\label{inhel}
\left( - \Box + m^2\right) \Lambda(x,y;m) =  \frac{1}{\sqrt{|g|}} \delta^{(d+1)}(x-y),
\end{equation}
where $m$ is the mass of the scalar field. Making use of the fact that $\Lambda$ is a function of the geodesic distance (or equivalently $K$), we find that
\begin{equation}
\label{boxopdef}
\Box  =   a^2 K^2 (1-K^2) \frac{\partial^2}{\partial K^2 }-  a^2 K \left(d-1+2 K^2\right)\frac{\partial}{\partial K } .
\end{equation}
The properly normalized solution to (\ref{inhel}) is well-known (e.g., ref. \cite{BURGES1986285}) and reads
\begin{equation}
\label{FullPropGennu}
\Lambda(K;m) = \frac{a^{d-1}\Gamma[\frac{\Delta}{2}] \Gamma[\frac{\Delta + 1}{2}]}{4\pi^{\frac{d+1}{2}} \Gamma[\Delta + 1 - \frac{d}{2}]} K^{\Delta}  ~{}_2F_1\left[\frac{\Delta}{2},\frac{\Delta + 1}{2}; \Delta + 1 - \frac{d}{2}; K^2\right]\mkern-3mu,
\end{equation} 
where $\Delta$ corresponds to the conformal dimension of the dual operator:
\begin{align}
\label{DeltaDef}
    \Delta = \frac{d}{2} \pm \sqrt{\frac{d^2}{4} + \frac{m^2}{a^2}}\,.
\end{align}
Bearing in mind the above relation, let us henceforth express the propagator by $\Lambda(K;\Delta)$, i.e., as a function of $K$ and $\Delta$. In the flat space limit and for $d>1$, $\Lambda(K;\Delta)$ reduces (up to a sign) to the Green's function of the Laplacian in $\mathbb{R}^{d+1}$:
\begin{equation}
\Lambda(K; \Delta) = \frac{\Gamma\left(\frac{d+1}{2} \right)}{2 (d-1) \pi^\frac{d+1}{2} r^{d-1}} + \mathcal{O}(a).
\end{equation} 

Our model is a conformally coupled scalar field with a quartic self--interaction propagating on a static $\mathbb{H}_4$ background, described by the action
\begin{equation}
\label{actionwithint}
S= \int \mathrm{d}^4x \sqrt{|g|} \left(\frac12 (\partial\phi)^2  + \frac{R}{12} \phi^2 + \lambda \frac{\phi^4}{4!}\right)\mkern-3mu,
\end{equation} 
where $\lambda>0$ is a dimensionless coupling constant and $R= - 12 a^2$ is the Ricci scalar.
Therefore, in what follows we consider $m^2 =-2a^2$, which corresponds to the dimensions $\Delta=1, 2$. Then, the propagator simplifies to
\begin{equation}
\label{prop}
\Lambda(K;\Delta) =\frac{a^2 K^\Delta}{4\pi^2 (1-K^2)}\qquad \mathrm{for}~\Delta=1,2.
\end{equation} 

Before closing this section let us express  schematically the expansions for the two- and the four-point function respectively, 
\begin{align}
\label{2ptdiagrams}
\begin{split}
\diagramthree ,
\end{split}
\end{align}

\begin{align}
\label{4ptdiagrams}
\begin{split}
\diagramfour ,
\end{split}
\end{align}
where we adopted the following Feynman rules:
\begin{itemize}
\item each line \raisebox{-0.1cm}{$\hbox{ \convertMPtoPDF{./diagrams/diagram1.0}{0.8}{0.8} }$} corresponds to the free two-point function  $\Lambda(x,y;\Delta)$,
\item each four-vertex \raisebox{-0.1cm}{$\hbox{ \convertMPtoPDF{./diagrams/diagram2.1}{0.8}{0.8} }$} stands for an integral $\int \mathrm{d}^4 x \sqrt{|g|}$ over the vertex point $x$.
\end{itemize}
One of the purposes of this work is to calculate all the above diagrams. In the following two sections, \ref{2ptfnchapter} and \ref{4ptfnchapter}, we will  focus on the one-particle irreducible diagrams, since all the above diagrams are simple products and/or permutations of the latter. At the beginning of each of these sections, there will be a figure displaying all the computed diagrams in the given section. The reader not interested in the details of the calculation can skip to section \ref{chapope}, where a summary of the obtained results is available.
\
\section{Two-point function}
\label{2ptfnchapter}

In this section, we compute the one-particle irreducible diagrams that contribute to the two-point function (\ref{2ptdiagrams}), or explicitly
\begin{align}
\label{1PI2pt}
\begin{split}
\diagramone
\end{split}.
\end{align}
As a warm-up we compute the mass shift diagram $\mathcal{I}_2$, and proceed afterwards in the calculation of the tadpole diagram $\mathcal{H}_2$ and the double tadpole diagram $\mathcal{L}_2$. Eventually, we discuss the technically more challenging sunset diagram $\mathcal{K}_2$. For $\Delta=1$ we will encounter infrared divergences which are absent for $\Delta=2$.

\subsection{The mass shift diagram}
\label{massshiftsection}
Strictly speaking, this diagram is not needed in our analysis. Nevertheless, it will appear as a counterterm of other diagrams. It is therefore convenient to have its form in order to identify such terms.  The mass shift diagram $\mathcal{I}_2$ depicted in figure (\ref{1PI2pt}) corresponds to the following integral
\begin{equation}
\label{massshiftbeginning}
\mathcal{I}_2 = \int \mathrm{d}^4x \sqrt{|g(x)|} \Lambda(x_1,x;\Delta) \Lambda(x_2,x;\Delta).
\end{equation}
Let us for the moment set the two points $x_1,x_2$ to $x_1 = (z_1,0), x_2 = ( z_2,0)$ (the covariant form will be restored later). By further denoting $x^2 \equiv {x^{i}}^2$, the integral is then
\begin{equation}
\label{I2int}
\mathcal{I}_2 = \frac{(4 z_1 z_2)^\Delta}{(4\pi^2)^2} \int \mathrm{d}^4x \frac{z^{2\Delta-4} (x^2 + z^2 + z_1^2)^{2-\Delta}(x^2 + z^2 + z_2^2)^{2-\Delta}}{\Pi_{i=1}^2 [x^2 + (z + z_i)^2][x^2 + (z - z_i)^2]},
\end{equation}
where we have used that $\sqrt{|g(x)|}=\frac{1}{a^4z^4}$. As already mentioned, the above integral features an IR divergence for $\Delta=1$, but not for $\Delta=2$. To continue, we introduce a dimensionless regulator $\sigma > 0$ which does not affect the $\Delta =2$ case:
\begin{equation}
\label{I2intregularized}
\mathcal{I}_2 = \frac{(4 z_1 z_2)^\Delta}{(4\pi^2)^2} \int \mathrm{d}^4x \frac{(z^{2} + \sigma^2 f^2(z_1,z_2))^{\Delta-2} (x^2 + z^2 + z_1^2)^{2-\Delta}(x^2 + z^2 + z_2^2)^{2-\Delta}}{\Pi_{i=1}^2 [x^2 + (z + z_i)^2][x^2 + (z - z_i)^2]},
\end{equation}
where $f(z_1,z_2)$ is a nonnegative function which will be determined later by imposing covariance.
\paragraph{$\mathbf{\Delta=1}$.}
\label{massubnu1}
The integral (\ref{I2intregularized}) can be integrated directly. Performing the $z$-integral and then integrating over $x$ using three-dimensional spherical coordinates yields in the limit of small $\sigma$

\begin{equation}
\label{I2indifficultcoordinates}
\mathcal{I}_2 = \frac{z_1 z_2 (z_1+z_2)^{-1}}{4 \pi  \sigma f(z_1,z_2) } + \frac{ z_1 z_2}{4 \pi^2 (z_1^2 -z_2^2)^2} \left[ (z_1^2 + z_2^2) \log \frac{16 z_1^2 z_2^2}{(z_1^2 -z_2^2)^2} + 2z_1 z_2 \log \frac{(z_1 -z_2)^2}{(z_1 + z_2)^2} \right].
\end{equation}
Then, the covariant form of the IR-finite contribution reads
\begin{equation}
\mathcal{I}^{\text{regular}}_2 = \frac{1}{8\pi^2} \left[ \frac{K}{1-K^2} \log 4 + \frac{K}{1-K^2} \log \frac{K^2}{1-K^2} + \frac{K^2}{1-K^2}\log \frac{1-K}{1+K}\right].
\end{equation}
On the other hand, the IR-divergent term is not generally covariant for a generic choice of $f(z_1,z_2)$. To determine this function we note that a covariant regularization can be obtained by continuation in $\Delta$, for instance. However, for noninteger values of $\Delta$ the scalar propagator on AdS is complicated. Still we can proceed, using the fact that for any covariant infrared regularization, $\mathcal{I}_2$ has to obey the inhomogenenous differential equation
\begin{equation}
\label{diffeqmasshift}
 (-\Box_{x_1} - 2 a^2) ~\mathcal{I}_2 = \Lambda (x_1,x_2;\Delta).
\end{equation}
For $\Delta=1$, the most general covariant solution is 
\begin{equation}
\label{I2ingeneralcovariantway}
\mathcal{I}_2 = c_1 \frac{K}{1-K^2} + c_2 \frac{K^2}{1-K^2} +  \frac{1}{8\pi^2} \left[\frac{K}{1-K^2} \log \frac{K^2}{1-K^2} + \frac{K^2}{1-K^2}\log \frac{1-K}{1+K}\right]\mkern-3mu,
\end{equation}
where the homogeneous part with constants $c_1,c_2$ corresponds to free $\Delta=1,2$ propagators. Now consider the boundary limit:
\begin{equation}
\mathcal{I}_2 \sim \left( c_1 + \frac{\log K}{4\pi^2}\right)K + c_2 K^2 + \mathcal{O}(K^3).
\end{equation}
The divergent term, whenever an AdS invariant IR regulator is available, should enter in $c_1$ or $c_2$. The term proportional to $c_2$ produces a fall-off behaviour corresponding to the $\Delta=2$ boundary condition. Thus, for $\Delta=1$, we set $c_2=0$. The divergent part should therefore be parameterized by $c_1$. Comparing eq. (\ref{I2ingeneralcovariantway}) with eq. (\ref{I2indifficultcoordinates}) uniquely fixes $f(z_1,z_2)$ up to a scale as 
\begin{equation}
\label{fforcov}
    f(z_1,z_2) = \pi \frac{(z_1-z_2)^2 (z_1 + z_2)}{z_1^2 + z_2^2}.
\end{equation}
The covariant form of eq. (\ref{I2indifficultcoordinates}) then reads
\begin{equation}\notag
\mathcal{I}_2 = \frac{1}{8\pi^2} \left[ \frac{K}{1-K^2} \left(\log 4 + \frac{1}{{\sigma}}\right) + \frac{K}{1-K^2} \log \frac{K^2}{1-K^2} + \frac{K^2}{1-K^2}\log \frac{1-K}{1+K}\right]\mkern-3mu,
\end{equation}
with boundary limit
\begin{equation}
\label{massnu1}
\mathcal{I}_2 \sim I_2=\frac{K}{8 \pi^2 \sigma} + \frac{ \log 2 K}{4\pi^2} K + \mathcal{O}(K^3).
\end{equation}

\paragraph{$\mathbf{\Delta=2}$.}

For $\Delta=2$, where there are no IR issues, the evaluation of the integral (\ref{I2int}) can again be carried out straightforwardly:
\begin{equation}
\mathcal{I}_2 = -\frac{2 z_1 z_2}{8 \pi^2 (z_1^2 -z_2^2)^2} \left[ (z_1^2 + z_2^2) \log \frac{(z_1 -z_2)^2}{(z_1 + z_2)^2} +2  z_1 z_2 \log \frac{16 z_1^2 z_2^2}{(z_1^2 -z_2^2)^2} \right]\mkern-4mu,
\end{equation}
which corresponds to the covariant expression
\begin{equation}
\mathcal{I}_2 = - \frac{1}{8\pi^2} \left[ \frac{K^2}{1-K^2} \log 4 + \frac{K}{1-K^2} \log \frac{1-K}{1+K} + \frac{K^2}{1-K^2}\log\frac{K^2}{1-K^2}\right]\mkern-4mu.
\end{equation}
This solution features two properties. Firstly, it solves as well the differential equation (\ref{diffeqmasshift}). Secondly, its behavior at the boundary exhibits a clear similarity to the regular part of the same expression for $\Delta=1$ (cf. eq. (\ref{massnu1})):
\begin{equation}
\label{massnu2}
\mathcal{I}_2 \sim I_2= \frac{1 - \log 2K}{4\pi^2}K^2  + \mathcal{O}(K^3).
\end{equation}
As expected, the leading terms of both eqs. \eqref{massnu1}, \eqref{massnu2} are of order $K^\Delta$.

\subsection{The tadpole diagram}
\label{tadpolesection}
The tadpole diagram $\mathcal{H}_2$ given in figure (\ref{1PI2pt}) has the integral expression
\begin{equation}
\label{tadpolediag}
\mathcal{H}_2 = \int \mathrm{d}^4x \sqrt{|g(x)|} \Lambda(x,x_1;\Delta) \Lambda(x,x_2;\Delta)\Lambda(x,x;\Delta).
\end{equation}
The expression $\Lambda(x,x;\Delta)$ is clearly ultraviolet divergent, and needs to be regularized. In principle, we can set it to zero by hand, but we would like to introduce the regulator that we systematically use later on. In position space, UV divergences result in the limit of colliding points, where $K\rightarrow 1$. The following ``rescaling" is AdS-invariant and resolves the short distance singularity of $1/(1-K)$-like expressions
\begin{equation}
\label{KtoKepsilon}
    K \rightarrow \frac{K}{1+\epsilon}.
\end{equation}
With the help of eq. (\ref{flatspacelimitofK}), we find that in the flat space limit the $\epsilon$ regularization takes the form 
\begin{equation}
1-\frac{K}{1 + \epsilon} = \frac{\frac{a^2r^2}{2} + \epsilon}{1 + \epsilon} + \mathcal{O}(a^4).
\end{equation}
From this it becomes clear that the above regularization carves out a small $\epsilon$-ball around the point and then rescales it by $1/(1+\epsilon)$. This regularization procedure will be used systematically on every UV-divergent integral encountered in this work. 
More precisely, for propagators representing internal lines, we rescale each $K$ as in eq. (\ref{KtoKepsilon}), i.e.,
\begin{equation}
\label{regularization}
\Lambda(x,y;\Delta) \rightarrow (1+\epsilon)^{2-\Delta } \frac{a^2 K_{xy}^\Delta}{4\pi^2 (1+ K_{xy}+\epsilon)(1 - K_{xy}+\epsilon)},
\end{equation}
while for propagators representing external legs, only those $K$'s appearing in the numerator are rescaled. 
Here $K_{xy}$ stands for $K$ as a function of the points $x$ and $y$. 

Returning to the tadpole diagram, the propagator at coincident points is given by
\begin{equation}
\Lambda(x,x;\Delta) = \frac{ a^2 }{4\pi^2}\frac{(1+\epsilon)^{2-\Delta}}{\epsilon (2+\epsilon)}.
\end{equation}
Therefore, the tadpole diagram reduces to the mass-shift diagram (\ref{massshiftbeginning}) times a divergent prefactor
\begin{equation}
\label{curlHcomputed}
\mathcal{H}_2 = \frac{a^2 }{8\pi^2}\left(\frac{1}{\epsilon} + \frac{3}{2} - 3 \Delta \right)\mathcal{I}_2 + \mathcal{O}(\epsilon).
\end{equation}
After sending $x_1$ and $x_2$ to the boundary, eq. (\ref{curlHcomputed}) simplifies to
\begin{equation}
\mathcal{H}_2 \sim H_2 = \frac{a^2 }{ 8 \pi^2}\left(\frac{1}{\epsilon} + \frac{3}{2} - 3 \Delta \right) I_2 + \mathcal{O}(\epsilon),
\end{equation}
where $I_2$ is given in eq. \eqref{massnu1} and eq. \eqref{massnu2}. 

\subsection{The double tadpole diagram}
The double tadpole diagram $\mathcal{L}_2$ in figure (\ref{1PI2pt}) corresponds to the following integral
\begin{equation}
\label{dtdint}
    \mathcal{L}_2 = \int \mathrm{d}^4 x \sqrt{|g(x)|} \int \mathrm{d}^4 y \sqrt{|g(y)|} \Lambda(x_1,x;\Delta) \Lambda(x_2,x;\Delta) \Lambda(x,y;\Delta)^2 \Lambda(y,y;\Delta),
\end{equation}
which, containing two loops, requires again a regularization. Adopting the regularization described in section \ref{tadpolesection}, $ \mathcal{L}_2$ takes the form
\begin{equation}\notag
    \mathcal{L}_2 = \frac{ a^2 }{(4\pi^2)^3}\frac{(1+\epsilon)^{6-5\Delta}}{\epsilon (2+\epsilon)} \int \mathrm{d}^4 x \sqrt{|g(x)|} \Lambda(x_1,x;\Delta) \Lambda(x_2,x;\Delta) \int \frac{\mathrm{d}^4 y}{w^4}   \frac{K_{xy}^{2\Delta}}{[(1+\epsilon)^2-K_{xy}^2 ]^2}.
\end{equation}
Let us first consider the integral over $y$. By the substitution $(w,y^i) \rightarrow (w ,y^i + x^i)$, it displays manifest independence on the ``nonradial" coordinates $x^i$. Then, a simple rescaling argument can be used to show that the integral also does not depend on the radial coordinate $z$. Indeed, any rescaling $z \rightarrow \theta z,~\theta >0$ can be undone by a substitution of the form $y^i \rightarrow \theta y^i,~ w \rightarrow \theta w$. In particular, we can set $z$ to any value $z_0>0$.\footnote{We might as well set $z_0=1$, but since the integral has to be regularized for $\Delta=1$, we keep $z_0$ and show that the scaling symmetry survives regularization.} Thus, the nested integral  (\ref{dtdint}) factorizes as 
\begin{equation}
    \mathcal{L}_2 = \frac{ a^2 }{(4\pi^2)^3}\frac{(1+\epsilon)^{6-5\Delta}}{\epsilon (2+\epsilon)}~\mathcal{M}_2\times \mathcal{I}_2 \,,
\end{equation}
where $\mathcal{I}_2$ is the mass shift computed in section \ref{massshiftsection} and where
\begin{equation}
\label{dtpm2}
    \mathcal{M}_2 = \int \frac{\mathrm{d}^4 y}{w^4} \frac{K_{x_0y}^{2\Delta}}{[(1+\epsilon)^2-K_{x_0y}^2 ]^2}= (2 z_0)^{2\Delta}  \int_{-\infty}^{\infty} \frac{\mathrm{d}^4 y}{w^{4-2\Delta}} \frac{Q^{4-2\Delta} [y^2 +(w -z_0)^2 + \epsilon Q ]^{-2}}{[y^2 +(w +z_0)^2 + \epsilon Q ]^2}.
\end{equation}
In the above formula we defined $Q= y^2 + w^2 +z_0^2$. For $\Delta=1$, $\mathcal{M}_2$ exhibits another IR divergence. In order to introduce the same IR-regulator as in section \ref{massshiftsection}, note that the integral above is essentially the mass shift diagram (\ref{massshiftbeginning}) evaluated by setting $x_1 = x_2 = (z_0,0)$ after rescaling $K$ as in (\ref{KtoKepsilon}) to regulate the resulting UV divergence. It is not hard to see that $f(z_1,z_2)$ given in eq. (\ref{fforcov}) generalizes to
\begin{equation}
    f(z_1,z_2) = \pi \frac{[(z_1-z_2)^2 +\epsilon(z_1^2+z_2^2)] [(z_1 + z_2)^2+ \epsilon(z_1^2+z_2^2)]}{(z_1^2 + z_2^2)(z_1+z_2)},
\end{equation}
which, by colliding the points, reduces to $z_0$ up to a coefficient. Thus, we regulate the integral as follows: 
\begin{equation}\notag
    \mathcal{M}_2 = (2 z_0)^{2\Delta} \int\frac{\mathrm{d}^4 y}{(w^2 + \sigma^2 z_0^2)^{2-\Delta}} \frac{Q^{4-2\Delta}}{[y^2 +(w + z_0)^2 + \epsilon Q ]^2[y^2 +(w -z_0)^2 + \epsilon Q ]^2},
\end{equation}
where the coefficient in front of $\sigma$ is irrelevant due to the scale invariance in $z_0$ (which survives the regularization) explained above. 
Owing to the $w \rightarrow -w$ symmetry of the integral, let us double the integration domain:
\begin{equation}\notag
    \mathcal{M}_2 = \frac{(2 z_0)^{2\Delta}}{2}  \int_{-\infty}^{\infty} \frac{\mathrm{d}^4 y}{(w^2 + \sigma^2 z_0^2)^{2-\Delta}} \frac{Q^{4-2\Delta}}{[y^2 +(w + z_0)^2 + \epsilon Q ]^2[y^2 +(w -z_0)^2 + \epsilon Q ]^2}.
\end{equation}
\paragraph{$\mathbf{\Delta=1}$.}
For $\Delta=1$, the integral (\ref{dtpm2}) becomes
\begin{equation}
    \mathcal{M}_2 = 2 z_0^2  \int_{-\infty}^{\infty} \frac{\mathrm{d}^4 y}{w^2+\sigma^2 z_0^2} \frac{Q^2}{[y^2 +(w +z_0)^2 + \epsilon Q ]^2[y^2 +(w -z_0)^2 + \epsilon Q ]^2},
\end{equation}
or equivalently $\mathcal{M}_2 = \mathcal{M}_2^+ +\mathcal{M}_2^-$, where
\begin{equation}
\label{Mcurlpm}
    \mathcal{M}_2^\pm = \frac{z_0^2}{(1+\epsilon)^2}  \int_{-\infty}^{\infty} \frac{\mathrm{d}^4 y}{w^2+\sigma^2 z_0^2} \frac{1}{[y^2 +(w +z_0)^2 + \epsilon Q ][y^2 +(w \pm z_0)^2 + \epsilon Q ]}.
\end{equation}
In the last step we used the $w \rightarrow -w$ symmetry and the identity
\begin{equation}\notag
    2(1+\epsilon)Q\frac{ [y^2 +(w -z_0)^2 + \epsilon Q ]^{-1}}{[y^2 +(w +z_0)^2 + \epsilon Q ]} =  \frac{1}{[y^2 +(w +z_0)^2 + \epsilon Q ]} + \frac{1}{[y^2 +(w -z_0)^2 + \epsilon Q ]} .
\end{equation}
The easiest way to deal with integrals of this form is to implement Schwinger parameters. Introducing one Schwinger parameter for each of the three factors in the denominator, the integral (\ref{Mcurlpm}) reads
\begin{equation}
    \mathcal{M}_2^\pm = \frac{z_0^2}{(1+\epsilon)^2}  \int_{-\infty}^{\infty} \mathrm{d}^4 y \int^{\infty}_0 \mathrm{d} t_1 \mathrm{d} t_2 \mathrm{d} t_3~ e^{-(t_1+t_2)(1+\epsilon)Q - (t_1 \pm t_2) 2w z_0 - t_3 (w^2+\sigma^2 z_0^2)}.
\end{equation}
Now, the spatial integral is a straightforward Gaussian integral resulting in

\begin{equation}
\label{M2pmintermediate}
    \mathcal{M}_2^\pm = \frac{\pi^2 z_0^2}{(1+\epsilon)^{3}}  \int^{\infty}_0 \mathrm{d} t_1 \mathrm{d} t_2 \mathrm{d} t_3 \frac{e^{ -z_0^2 (1+\epsilon) \frac{(t_1+t_2+t_3)^2 - \frac{(t_1 \pm t_2)^2}{(1+\epsilon)^2} + (\sigma^2 -1) t_3 (t_1+t_2+t_3)} {t_1+t_2+t_3}}}{ (t_1 +t_2)^{\frac{3}{2}} \sqrt{t_1+t_2+t_3}},
\end{equation}
where we substituted $t_3 \rightarrow (1+\epsilon) t_3$. 
Let us introduce the following coordinates: $t_i = s s_i$, which simply correspond to a rescaling of our original coordinates $t_i$ by a factor $s$. Accordingly, one can rewrite the measure as
\begin{equation}
\Pi_{i=1}^n \mathrm{d}t_i =  \Pi_{i=1}^n \mathrm{d}s_i~ \int_{0}^{\infty}\mathrm{d}s ~s^{n-1}~\delta(1 - \sum_{i=1}^n s_i),\qquad \forall n \in \mathbb{N},
\end{equation}
 assuming that the condition $ s = \sum_i t_i $ is satisfied. Hence, eq. (\ref{M2pmintermediate}) becomes
 \begin{equation}
  \mathcal{M}_2^\pm = \frac{\pi^2 z_0^2}{(1+\epsilon)^{3}}  \int^{\infty}_0 \mathrm{d} s_1 \mathrm{d} s_2 \mathrm{d} s_3  \mathrm{d}s~\delta(1 - \sum_{i=1}^3 s_i)~\frac{e^{-s z_0^2 (1+\epsilon) \left(1 - \frac{(s_1 \pm s_2)^2}{(1+\epsilon)^2} + (\sigma^2 -1) s_3 \right)}}{ (s_1 +s_2)^{\frac{3}{2}}},
\end{equation}
where we already integrated over $s_3$. Integrating over $s$ yields
 \begin{equation}
  \mathcal{M}_2^\pm = \frac{\pi^2}{(1+\epsilon)^{4}}  \int^{\infty}_0 \mathrm{d} s_1  \mathrm{d} s_2 \mathrm{d} s_3 \frac{\delta (1- s_1 - s_2 - s_3)}{ (s_1 +s_2)^{\frac{3}{2}} \left[1 -\frac{(s_1 \pm s_2)^2}{(1+\epsilon)^2} + (\sigma^2 -1) s_3\right]},
\end{equation}
which further leads, in the limit $\sigma \rightarrow 0$, to
\begin{equation}
    \mathcal{M}_2^+ = \frac{\pi^2 }{(1+\epsilon)^4} \left(\frac{\pi}{\sigma} - 2 + 2 \frac{\mathrm{arccoth}(1+\epsilon) }{1+\epsilon}  \right) + \mathcal{O}(\sigma)
\end{equation}
and
\begin{equation}
    \mathcal{M}_2^- = \frac{\pi^2 }{(1+\epsilon)^4} \left(\frac{\pi}{\sigma} - 1 + \frac{\epsilon (2+\epsilon) \log \frac{\epsilon}{2+\epsilon}}{2(1+\epsilon)}  \right) + \mathcal{O}(\sigma). 
\end{equation}
Eventually, the full double tadpole diagram is given by
\begin{equation}
\label{finalL2delta2}
    \mathcal{L}_2 = \frac{ a^2 \pi^2 }{2 (4\pi^2)^3} \left(\frac{-3 +\frac{2\pi}{\sigma} - \log\frac{\epsilon}{2}}{\epsilon } +11 - \frac{7 \pi}{\sigma}  +\frac{11}{2} \log\frac{ \epsilon}{2}\right) \mathcal{I}_2+\mathcal{O}(\epsilon,\sigma).
\end{equation}

\paragraph{$\mathbf{\Delta=2}$.}
By setting $\Delta=2$ and rescaling $z_0$ to $1$, the integral (\ref{dtpm2}) becomes
\begin{equation}
    \mathcal{M}_2 = 8  \int_{-\infty}^{\infty} \mathrm{d}^4 y \frac{1}{[y^2 +(w +1)^2 + \epsilon Q ]^2[y^2 +(w -1)^2 + \epsilon Q ]^2}.
\end{equation}
Again, it is favourable to introduce Schwinger parameters. This results in
\begin{equation}
    \mathcal{M}_2 = 8 \int_{-\infty}^{\infty} \mathrm{d}^4 y \int_0^{\infty} \mathrm{d}t_1 \mathrm{d}t_2 ~t_1 t_2~ e^{-(t_1+t_2) (1+\epsilon) Q- (t_1-t_2) 2w},
\end{equation}
which allows for a simple spatial integration, yielding
\begin{equation}
    \mathcal{M}_2 = 8  \pi^2 \int_0^{\infty} \mathrm{d}t_1 \mathrm{d}t_2~\frac{ t_1 t_2}{(t_1 + t_2)^2} ~e^{\frac{(t_1 - t_2)^2}{t_1+t_2} - (t_1+t_2)(1+\epsilon)^2 }.
\end{equation}
After the substitution $t_i \rightarrow s s_i$, the integral becomes
\begin{equation}
    \mathcal{M}_2 = 8 \pi^2 \int_0^{\infty} \mathrm{d}s_1 \mathrm{d}s_2 \mathrm{d}s ~\frac{s~ s_1 s_2~\delta(1-s_1-s_2)}{(s_1+s_2)^2} ~e^{-s \frac{ (s_1+s_2)^2(1+\epsilon)^2-(s_1-s_2)^2}{s_1+s_2}} .
\end{equation}
 Then, by removing the Schwinger parameter $s$ and subsequently integrating over $s_2$, one gets
\begin{equation}
    \mathcal{M}_2 = 8 \pi^2 \int_0^{1} \mathrm{d}s_1 ~ ~\frac{s_1(1-s_1)}{[(1+\epsilon)^2-(2s_1-1)^2]^2}.
\end{equation}
Eventually, the evaluation of the last integral yields
\begin{equation}
    \mathcal{M}_2 = - \pi^2\frac{ 2 (1+\epsilon) + [2+\epsilon  (2+\epsilon)] \log \frac{\epsilon}{2+\epsilon}}{2 (1+\epsilon)^3}.
\end{equation}
For small $\epsilon$, the full solution for the double tadpole diagram reads
\begin{equation}
    \mathcal{L}_2 = \frac{ a^2 \pi^2 }{2(4\pi^2)^3} \left( \frac{14+13 \log \frac{\epsilon}{2}}{2}-\frac{1+\log\frac{\epsilon}{2}}{\epsilon } \right) \mathcal{I}_2+\mathcal{O}(\epsilon).
\end{equation}
We close this section with a comment on the renormalization scale. When analyzing QFT in AdS typically one encounters a separation of scales into a UV scale which is related to the scale of local physics and an IR scale given by the AdS radius. In the present context, however, we will be interested in boundary-to-boundary correlation functions for which the AdS radius is the only relevant scale and the UV scale is absent. This is also implicit in the choice of the dimensionless regulator $\epsilon$ in eq. (\ref{KtoKepsilon}), which is related to a dimensionful cut-off $\Lambda$ through $\Lambda=\epsilon/a$. 

\subsection{The sunset diagram}\label{ssd}
The sunset diagram $\mathcal{K}_2$ in figure (\ref{1PI2pt}) instructs us to compute
\begin{equation}
\label{mathcalK2formula}
\mathcal{K}_2 =    \int \mathrm{d}^4 y \sqrt{|g(y)|}\int \mathrm{d}^4 x \sqrt{|g(x)|} \Lambda(x_1,x;\Delta) \Lambda(x,y;\Delta)^3  \Lambda(x_2,y;\Delta) .
\end{equation}
Let us split it into two parts, the first one being
\begin{equation}
\label{jellyfishbeginning}
\mathcal{J}_2 = \int \mathrm{d}^4 y  \sqrt{|g(y)|} \Lambda (x,y;\Delta)^3 \Lambda (x_2,y;\Delta).
\end{equation}
As we are eventually interested in the anomalous dimensions of the operators on the boundary we can already take the boundary limit for $x_2$. Then, $\mathcal{J}_2 $ takes the following form
\begin{equation}
\mathcal{J}_2 \sim J_2 = (1+\epsilon)^{6-5\Delta}\frac{a^8 z_2^\Delta}{(4\pi^2)^4} \int \mathrm{d}^4 y  \sqrt{|g(y)|} \frac{K_{xy}^{3\Delta}}{(1-K_{xy}+\epsilon)^3 (1+K_{xy}+\epsilon)^3} \bar{K}_{x_2 y}^\Delta\,.
\end{equation}
In the last step we also introduced the UV regulator, cf. eq. (\ref{regularization}).
Using translation symmetry, we can shift $x$ and $x_2$ by $(0,- x_2^i)$, which leads to
\begin{equation}
J_2 =(1+\epsilon)^{6-5\Delta}\frac{a^8 z_2^\Delta}{(4\pi^2)^4} \int \mathrm{d}^4 y  \sqrt{|g(y)|} \frac{K_{x' y}^{3\Delta}}{(1-K_{x'y}+\epsilon)^3 (1+K_{x'y}+\epsilon)^3} \bar{K}_{x_2' y}^\Delta\,,
\end{equation}
where $x_2'=(0, 0)$. Next, as was done in ref. \cite{Freedman:1998tz}, we use inversion symmetry\footnote{The inversion operator $I$ acts on conformal coordinates as $I(x^i) =  \frac{x^i}{x^{2}+z^2},~ I(z) = \frac{z}{x^{2} + z^2}$.} to simplify the above expression. In particular, we invert every point by itself, which therefore results in sending $x'_2$ to infinity. For the propagators, this gives
\begin{equation}
K_{x ' y} = K_{x'' y''},\qquad  \bar{K}_{x_2' y} = x_2''^2 \bar{K}_{x_2'' y''} = 2w'',
\end{equation}
where we denoted the inverted points by double primes. Note that the measure of the integral does not change under the inversion. A subsequent variable substitution $(w'',y^{i''}) = (w, y^i + x^{i''}) $ eventually gives
\begin{equation}
\label{J2nu}
J_2 = (1+\epsilon)^{6-5\Delta} \frac{a^4 (16 z_2)^\Delta z''^{3\Delta}}{2 (4\pi^2)^4} \int_{-\infty}^{\infty} \mathrm{d}^3 y ~ \mathrm{d}w ~ \frac{ Q^{6-3\Delta}[y^2 + (z''-w)^2 + \epsilon Q]^{-3}}{w^{4-4\Delta}[y^2 + (z''+w)^2+\epsilon Q]^3},
\end{equation}
where $Q = y^2 + z''^2 + w^2$ and we used again the fact that the integrand is symmetric under $w \rightarrow - w$.

\paragraph{$\mathbf{\Delta=1}$.}
For $\Delta=1$, the integral (\ref{J2nu}) becomes
\begin{equation}\notag
J_2 =(1+\epsilon) \frac{8 a^4  z_2 z''^{3}}{ (4\pi^2)^4}\int_{-\infty}^{\infty} \mathrm{d}^3 y ~ \mathrm{d}w ~ \frac{   Q^{3}}{[(y^2 + (z''-w)^2 +  \epsilon Q)(y^2 + (z''+w)^2+  \epsilon Q)]^3}.
\end{equation}
A decomposition in partial fractions yields
\begin{equation}\notag
J_2 =  \frac{a^4  z_2 z''^{3}}{ (4\pi^2)^4(1+\epsilon)^2}\int_{-\infty}^{\infty} \mathrm{d}^3 y ~ \mathrm{d}w ~ \left[\frac{1}{y^2 + (z''-w)^2 +  \epsilon Q} + \frac{1}{y^2 + (z''+w)^2+  \epsilon Q}\right]^3\mkern-5mu.
\end{equation}
Using the $w \rightarrow - w$ symmetry, one can write $J_2 =J_2^a +J_2^b$, where
\begin{equation}
J_2^a =  \frac{2 a^4  z_2 z''^{3}}{ (4\pi^2)^4(1+\epsilon)^2}\int_{-\infty}^{\infty} \mathrm{d}^3 y  ~ \mathrm{d}w ~  \frac{1}{ [(1+\epsilon)Q + 2 z'' w]^3} 
\end{equation}
and
\begin{equation}
J_2^b =  \frac{6 a^4 z_2 z''^{3}}{ (4\pi^2)^4(1+\epsilon)^2}\int_{-\infty}^{\infty} \mathrm{d}^3 y  ~ \mathrm{d}w ~  \frac{1}{[(1+\epsilon)Q + 2 z'' w]^2 [(1+\epsilon)Q - 2 z'' w]}.
\end{equation}
The computation of $J_2^a$ is rather simple.
Indeed, the introduction of one Schwinger parameter gives
\begin{equation}
J_2^{a} = \frac{a^4  z_2 z''^{3}}{ (4\pi^2)^4(1+\epsilon)^2} \int_{0}^{\infty}\mathrm{d}t ~t^2 \int_{-\infty}^{\infty} \mathrm{d}^3 y ~ \mathrm{d}w ~ e^{- t(1+\epsilon) Q - 2 t z'' w},
\end{equation}
allowing for the integration over space,
\begin{equation}
J_2^{a} =  \frac{\pi^2  a^4  z_2 z''^{3}}{ (4\pi^2)^4(1+\epsilon)^4} \int_{0}^{\infty}\mathrm{d}t ~ e^{- t z''^2 (1+\epsilon) \left[1- \frac{1}{(1+\epsilon)^2}\right]} = \frac{a^4  z_2 z''}{ (4\pi^2)^4} \frac{\pi^2}{(1+\epsilon)^3\epsilon (2+\epsilon) }.
\end{equation}
For small values of $\epsilon$, this simplifies to
\begin{equation}
J_2^{a} =  \frac{\pi^2 a^4  z_2 z''}{ (4\pi^2)^4}\left(\frac{1}{2\epsilon}- \frac{7}{4} \right) + \mathcal{O}(\epsilon).
\end{equation}
For the calculation of $J_2^b$, the introduction of two Schwinger parameters is more convenient,
\begin{equation}
J_2^b = \frac{6 a^4 z_2 z''^{3}}{ (4\pi^2)^4(1+\epsilon)^2} \int_{0}^{\infty}\mathrm{d}t_1 \mathrm{d}t_2 ~t_1~\int_{-\infty}^{\infty} \mathrm{d}^3 y  ~ \mathrm{d}w ~  e^{-(t_1+t_2) (1+\epsilon)Q-(t_1-t_2) 2 z'' w}.
\end{equation}
Integration over space is now viable and results in
 \begin{equation}
J_2^b =  \frac{6  \pi^2  a^4 z_2 z''^{3}}{ (4\pi^2)^4(1+\epsilon)^4} \int_{0}^{\infty}\mathrm{d}t_1 \mathrm{d}t_2 ~\frac{t_1}{(t_1+t_2)^2}~ e^{-(t_1+t_2) (1+\epsilon) z''^2 \left[1 - \frac{(t_1-t_2)^2}{(t_1+t_2)^2(1+\epsilon)^2}\right]}.
\end{equation}
Note that the integral does not change if we swap $t_1$ and $t_2$. Therefore one can replace the $t_1$ in front of the exponential with $(t_1+t_2)/2$.  After introducing new coordinates $t_i = s s_i$ as done above, the integration over $s_2$ leads to
 \begin{equation}
J_2^b =   \frac{3 \pi^2  a^4 z_2 z''^{3}}{ (4\pi^2)^4(1+\epsilon)^4} \int_{0}^{\infty}\mathrm{d}s ~\int_{0}^{1}\mathrm{d}s_1 ~  e^{- s (1+\epsilon) z''^2 \left[1 - \frac{(1-2s_1)^2}{(1+\epsilon)^2}\right]}.
\end{equation}
Removing the Schwinger parameter $s$ allows one to integrate over $s_1$. Therefore, the final result for $J_2^b$ is given by
\begin{equation}
J_2^b =  \frac{3 \pi^2  a^4 z_2 z''}{ 2(4\pi^2)^4(1+\epsilon)^4}\log\frac{2+\epsilon}{\epsilon}= -  \frac{3 \pi^2  a^4 z_2 z''}{2 (4\pi^2)^4}\log \frac{\epsilon}{2} + \mathcal{O}(\epsilon).
\end{equation}

\paragraph{$\mathbf{\Delta=2}$.}
The integral (\ref{J2nu}) is instead given by
\begin{equation}\notag
J_2 =  \frac{a^4 (16 z_2)^2 z''^{6}}{2 (4\pi^2)^4 (1+\epsilon)^{4}}\int_{-\infty}^{\infty} \mathrm{d}^3 y ~ \mathrm{d}w ~ \frac{w^4}{[y^2 + (z''+w)^2 + \epsilon Q]^3[y^2 + (z''-w)^2+\epsilon Q]^3}.
\end{equation}
As usual, let us introduce the Schwinger parameters
\begin{align}\notag
\begin{split}
J_2 =&\frac{a^4 (16 z_2)^2 z''^{6}}{8 (4\pi^2)^4 (1+\epsilon)^{4}}\int_{0}^{\infty}\mathrm{d}t_1 \mathrm{d}t_2 ~(t_1 t_2)^2~\int_{-\infty}^{\infty} \mathrm{d}^3 y  ~ \mathrm{d}w ~ w^4~ e^{-(t_1+t_2) (1+\epsilon)Q-(t_1-t_2) 2 z'' w}\\
=& \frac{a^4 (16 z_2)^2 z''^{6}\partial^2_\gamma |_{\gamma =1}}{8 (4\pi^2)^4 (1+\epsilon)^{6}} \int_{0}^{\infty}\frac{\mathrm{d}t_1 \mathrm{d}t_2~(t_1 t_2)^2}{(t_1+t_2)^2} \int_{-\infty}^{\infty} \mathrm{d}^3 y  ~ \mathrm{d}w~e^{-(t_1+t_2) (1+\epsilon)(y^2 + \gamma w^2 + z''^2)-(t_1-t_2) 2 z'' w},
\end{split}
\end{align}
where in the last step we introduced an auxiliary parameter $\gamma>0$ in order to get rid of the factor $w^4$ in the numerator. The spatial integration yields
\begin{equation}\notag
J_2 = \frac{a^4 \pi^2 (16 z_2)^2 z''^{6}}{8 (4\pi^2)^4 (1+\epsilon)^{8}}\partial^2_\gamma |_{\gamma =1}\frac{1}{\sqrt{\gamma}}\int_{0}^{\infty}\mathrm{d}t_1 \mathrm{d}t_2 ~\frac{(t_1 t_2)^2}{(t_1+t_2)^4}~e^{-(t_1+t_2) (1+\epsilon) z''^2 \left[1 - \frac{(t_1-t_2)^2}{(t_1+t_2)^2(1+\epsilon)^2\gamma}\right]}.
\end{equation}
Let us now apply the substitution $t_i = s s_i$. After integrating over $s_2$, we get
\begin{equation}\notag
J_2 = \frac{a^4 \pi^2 (16 z_2)^2 z''^{6}}{8 (4\pi^2)^4 (1+\epsilon)^{8}}\partial^2_\gamma |_{\gamma =1}\frac{1}{\sqrt{\gamma}} \int_{0}^{\infty}\mathrm{d}s ~\int_{0}^{1}\mathrm{d}s_1 ~s~ s_1^2 (1-s_1)^2  e^{- s (1+\epsilon) z''^2 \left[1 - \frac{(1-2s_1)^2}{(1+\epsilon)^2\gamma}\right]}.
\end{equation}
Integrating over the Schwinger parameter $s$ gives
\begin{equation}
J_2 = \frac{a^4 \pi^2 (16 z_2)^2 z''^{2}}{8 (4\pi^2)^4 (1+\epsilon)^{6}}\partial^2_\gamma |_{\gamma =1}~\gamma^{3/2} \int_{0}^{1}\mathrm{d}s_1~   \left[\frac{s_1 (1-s_1)}{ (1+\epsilon)^2 \gamma - (1-2s_1)^2}\right]^2\mkern-5mu.
\end{equation}
Differentiating twice by $\gamma$, setting $\gamma=1$, and subsequently integrating over $s_1$, yields
\begin{equation}
J_2 =  \frac{a^4 \pi^2  z_2^2 z''^{2}}{ (4\pi^2)^4} \left[\frac{1}{\epsilon} + \frac{1}{2} (-1+6 \log\frac{\epsilon}{2}) \right] +\mathcal{O}(\epsilon).
\end{equation}

\paragraph{Recovering the full covariance.}
In order to get back the explicit covariant form, let us first note that undoing the inversion and restoring the translation invariance instructs us to replace 
\begin{equation}
2 z'' \rightarrow \bar{K}_{x x_2},
\end{equation} 
which is manifestly covariant. Therefore, it follows that one can write $J_2$ as
\begin{equation}
J_2 = \frac{a^4  \pi^2}{4 (4 \pi^2)^4}   K^\Delta_{x x_2}\left(\frac{1}{\epsilon} +3 (-1)^\Delta \log \frac{\epsilon}{2} -\frac{13}{2} + 3\Delta \right)\mkern-3mu.
\end{equation}
Furthermore, due to the covariant form of $\mathcal{J}_2$ in eq. (\ref{jellyfishbeginning}), it has to correspond to the above result modulo some function of $K_{x x_2}$. Considering also the symmetry between $x_1$ and $x_2$ in eq. (\ref{mathcalK2formula}), one concludes
\begin{equation}
\mathcal{J}_2 = \frac{a^2  \pi^2}{4 (4 \pi^2)^3}  \Lambda(x,x_2; \Delta) \left(\frac{1}{\epsilon} +3 (-1)^\Delta \log \frac{\epsilon}{2} -\frac{13}{2} + 3\Delta \right)\mkern-3mu.
\end{equation}

\paragraph{Attaching the missing leg.}

The full sunset diagram can be obtained by attaching one more leg to $\mathcal{J}_2$ we just extracted the most singular part of, i.e.,
\begin{equation}
\mathcal{K}_2 = \int \mathrm{d}^4 x \sqrt{|g(x)|} \Lambda(x_1,x;\Delta)\mathcal{J}_2 .
\end{equation}
Then the final result can again be expressed in terms of the mass shift $\mathcal{I}_2$
\begin{equation}
\label{sunsetfull}
\mathcal{K}_2 = \frac{a^2  \pi^2}{4(4\pi^2)^3} \left(\frac{1}{\epsilon} +3 (-1)^\Delta \log \frac{\epsilon}{2} -\frac{13}{2} + 3\Delta \right) \mathcal{I}_2.
\end{equation}

\section{Four-point function} 
\label{4ptfnchapter}
In this section we compute the diagrams that contribute to the four-point function. Up to the second order in the coupling constant $\lambda$, the one-particle irreducible diagrams are
\begin{align}
\label{1PI4pt}
\begin{split}
\diagramtwo
\end{split}.
\end{align}
The contact cross diagram $\mathcal{I}_4$ is well-known in the literature, see, for instance, ref. \cite{Muck:1998rr}, and we just quote the result. Afterwards, we compute the one loop diagram $\mathcal{K}_4$.

\subsection{The cross diagram}
The cross diagram leads to
\begin{equation}
\mathcal{I}_4 = \int \mathrm{d}^4x \sqrt{|g(x)|} \Lambda(x_1,x;\Delta)\Lambda(x_2,x;\Delta)\Lambda(x_3,x;\Delta)\Lambda(x_4,x;\Delta),
\end{equation}
With all external legs on the boundary the integral reduces to
\begin{equation}
\mathcal{I}_4 \sim   \frac{4^{2\Delta} a^4 \Pi_{i=1}^4 z_i^\Delta}{(4\pi^2)^4} I_4,
\end{equation}
with
\begin{equation}
I_4 = \int \mathrm{d}^4 x \frac{z^{4\Delta-4}}{\Pi_{i=1}^4 \left[ (x-x_i)^2 + z^2\right]^\Delta}.
\end{equation}
As opposed to the rest of this paper, note the extra factor between ${\mathcal{I}}_4$ and $I_4$ which we introduced in order to conform with the literature. The result for $I_4$ is \cite{Muck:1998rr}
\begin{equation}
\label{finalformula}
I_4 = \frac{2 \pi^{\frac{3}{2}}\Gamma\left(2\Delta - \frac{3}{2}\right)}{ \Gamma(2\Delta)(\eta \zeta~ \Pi_{i<j} r_{ij})^{\frac{2}{3}\Delta}}  \int_{0}^{\infty} \mathrm{d}z ~{}_2F_1\left[\Delta,\Delta;2\Delta; 1 - \left(\frac{\eta  + \zeta }{\eta  \zeta }\right)^2 - \frac{4\sinh^2z}{\eta  \zeta}\right]\mkern-3mu,
\end{equation}
with the conformal invariants defined as
\begin{equation}
\label{confinvariantlist}
\eta = \frac{r_{14}r_{23}}{r_{12}r_{34}},\qquad \zeta = \frac{r_{14}r_{23}}{r_{13}r_{24}}.
\end{equation}
For our purposes a more convenient form for $I_4$ is given by \cite{Dolan:2000uw}
\begin{align}
\label{MellinI4}
\begin{split}
I_4 =&\frac{\pi^{\frac{3}{2}}\Gamma\left(2 \Delta - \frac{3}{2}\right)}{  \Gamma(\Delta)^4} \frac{v^\Delta}{ (r_{12}r_{34})^{2\Delta} }\sum_{m,n=0}^\infty \frac{Y^m v^n}{m! (n!)^2 } \frac{ \Gamma(n+ \Delta)^2 \Gamma(n+ m+ \Delta)^2}{\Gamma(2n+m+ 2\Delta)}  \\
&\times \left[\psi(n+1) -\frac12 \log v  - \psi(n+\Delta) - \psi(n+\Delta + m) + \psi(2n + 2\Delta +m)\right]\mkern-3mu,
\end{split}
\end{align}
where we introduced the new invariants
\begin{equation}
    Y = 1 - \frac{1}{\zeta^2},\qquad v = \frac{1}{\eta^2}.
\end{equation}

\subsection{The one loop diagram}
The one loop diagram $\mathcal{K}_4$ depicted in figure (\ref{1PI4pt}) is given by the double integral
\begin{equation}\label{K_4c}
\mathcal{K}_4 = \int \mathrm{d}^4x \sqrt{|g(x)|} \Lambda(x_1,x;\Delta)\Lambda(x_2,x;\Delta) \mathcal{J}_4,
\end{equation}
where we defined the sub--integral $\mathcal{J}_4$ as
\begin{equation}
\mathcal{J}_4 = \int \mathrm{d}^4 y  \sqrt{|g(y)|} \Lambda (x,y;\Delta)^2 \Lambda (x_3,y;\Delta)\Lambda (x_4,y;\Delta).
\end{equation}
The latter, by sending $z_3$ and $z_4$ to the boundary, takes the form
\begin{equation}\notag
\mathcal{J}_4 \sim J_4 = (1+\epsilon)^{4-6\Delta}\frac{a^8 ( z_3 z_4)^\Delta}{(4\pi^2)^4} \int \mathrm{d}^4 y  \sqrt{|g(y)|} \frac{K_{xy}^{2\Delta}}{(1-K_{xy}+\epsilon)^2 (1+K_{xy}+\epsilon)^2} \bar{K}_{x_3 y}^\Delta \bar{K}_{x_4 y}^\Delta.
\end{equation}
As before, the UV divergences are regularized with the help of $\epsilon$:  $K_{xy} \rightarrow K_{xy}/(1+\epsilon)$. Furthermore, since the divergence is logarithmic in $\epsilon$, we can safely ignore the prefactor $(1+\epsilon)^{4-6\Delta}$ in what follows. Translating the points $x, x_3$, and $x_4$ by $(0,- x^i_4)$ yields
\begin{equation}
J_4 =\frac{a^8 ( z_3 z_4)^\Delta}{(4\pi^2)^4} \int \mathrm{d}^4 y  \sqrt{|g(y)|} \frac{K_{x' y}^{2\Delta}}{(1-K_{x'y}+\epsilon)^2 (1+K_{x'y}+\epsilon)^2} \bar{K}_{x_3' y}^\Delta \bar{K}_{x_4' y}^\Delta\,,
\end{equation}
where $x_4'=(0, 0)$. As detailed in ref. \cite{Freedman:1998tz}, we use the inversion trick to simplify the above expression. For the $K$'s this gives
\begin{equation}
K_{x ' y} = K_{x'' y''},\qquad \bar{K}_{x_3' y} = \frac{1}{x_3'^2} \bar{K}_{x_3'' y''}=\frac{1}{r_{34}^2} \bar{K}_{x_3'' y''},\qquad \bar{K}_{x_4' y} = x_4''^2 \bar{K}_{x_4'' y''} = 2w'',
\end{equation}
where the inverted points are denoted by double primes. One more substitution $(w'',y^{i''}) = (w, y^i + x^{i''}) $ eventually yields
\begin{equation}
\label{J4nu}
J_4 = \frac{a^4 (16 z_3 z_4)^\Delta z''^{2\Delta}}{2 (4\pi^2)^4 r_{34}^{2\Delta}} \int_{-\infty}^{\infty} \mathrm{d}^3 y ~ \mathrm{d}w ~ \frac{Q^{4-2\Delta} w^{4\Delta-4}  [y^2 + (z''-w)^2 + \epsilon Q]^{-2}}{[(x_3'''-y)^2 + w^2]^{\Delta}[y^2 + (z''+w)^2+\epsilon Q]^2}\,,
\end{equation}
with $x_3^{i'''} = x_3^{i''} - x^{i''}$, $Q = y^2 + z''^2 + w^2$. Again, we used the symmetry of the integrand under $w \rightarrow - w$.

\paragraph{$\mathbf{\Delta=1}$.}
For this value, the integral (\ref{J4nu}) takes the form
\begin{equation}
J_4 = \frac{8 a^4  z_3 z_4 z''^{2}}{(4\pi^2)^4 r_{34}^{2}}\int_{-\infty}^{\infty} \mathrm{d}^3 y ~ \mathrm{d}w ~ \frac{   Q^{2}[(x_3'''-y)^2 + w^2]^{-1}}{[(y^2 + (z''-w)^2 +  \epsilon Q)(y^2 + (z''+w)^2+  \epsilon Q)]^2}.
\end{equation}
The $Q$ in the numerator can be written as 
\begin{align}
\begin{split}
 [2 (1+\epsilon) Q]^{2} =&[y^2 + (z'' - w)^2+ \epsilon Q]^{2} + [y^2 + (z'' + w)^2+ \epsilon Q]^{2}\\
 &+2 [y^2 + (z'' + w)^2+ \epsilon Q][y^2 + (z'' - w)^2+ \epsilon Q],
 \end{split}
\end{align}
which allows for a splitting of the integral in a divergent part and a regular part: $J_4 = J_4^d +J_4^r$. Explicitly, one gets
\begin{equation}
J_4^{d} = \frac{4 a^4  z_3 z_4 z''^{2}}{(4\pi^2)^4 r_{34}^{2}(1+\epsilon)^2} \int_{-\infty}^{\infty} \mathrm{d}^3 y ~ \mathrm{d}w ~ \frac{1}{[y^2 + (z''-w)^2+ \epsilon Q]^2[(x_3'''-y)^2 + w^2]}
\end{equation}
and
\begin{equation}
\label{regintintstep}
J_4^{r} =  \frac{4 a^4  z_3 z_4 z''^{2}}{(4\pi^2)^4 r_{34}^{2}} \int_{-\infty}^{\infty} \mathrm{d}^3 y ~ \mathrm{d}w ~ \frac{1}{[y^2 + (z''-w)^2][y^2 + (z''+w)^2][(x_3'''-y)^2 + w^2]},
\end{equation}
where for the latter integral, being regular, we already set $\epsilon \rightarrow 0$. Additionally, the regular integral can be further simplified to
\begin{equation}
J_4^{r} = \frac{4 a^4  z_3 z_4 z''^{2}}{(4\pi^2)^4 r_{34}^{2}} \int_{-\infty}^{\infty} \mathrm{d}^3 y ~ \mathrm{d}w ~ \frac{1}{[y^2 + (z''-w)^2][y^2 + z''^2+w^2][(x_3'''-y)^2 + w^2]},
\end{equation}
by using
\begin{equation}
\frac{2Q}{[y^2 + (z''-w)^2][y^2 + (z''+w)^2]} = \frac{1}{y^2 +(z'' - w)^2}+\frac{1}{y^2 +(z'' + w)^2}
\end{equation}
and the symmetry $w\rightarrow -w$ of the integrand of eq. (\ref{regintintstep}).
The computation of $J_4^d$ is rather simple. Introducing two Schwinger parameters and integrating over space yields
\begin{equation}
J_4^{d} =  \frac{4\pi^2 a^4  z_3 z_4 z''^{2}}{(4\pi^2)^4 r_{34}^{2}(1+\epsilon)^4} \int_{0}^{\infty}\mathrm{d}t_1\mathrm{d}t_2~ t_1 ~e^{- \frac{t_1 t_2 x_3'''^2}{t_1+t_2}} e^{ - \frac{t_1(t_2+ \frac{\epsilon}{1+\epsilon} t_1) z''^2}{(1+\epsilon)(t_1+t_2)}}~e^{-\frac{\epsilon}{1+\epsilon} t_1 z''^2}.
\end{equation}
After the change of variables $t_i = s s_i$, one then finds
\begin{align}\notag
J_4^{d} &=   \frac{4\pi^2 a^4  z_3 z_4 z''^{2}}{(4\pi^2)^4 r_{34}^{2}(1+\epsilon)^4} \int_{0}^{\infty}\mathrm{d}s_1\mathrm{d}s_2\mathrm{d}s~s_1\delta(1- s_1-s_2) ~ e^{-s \left[s_1 s_2 x_3'''^2 + \frac{s_1(s_2+ \frac{\epsilon}{1+\epsilon} s_1) z''^2}{(1+\epsilon)}+\frac{\epsilon}{1+\epsilon} s_1 z''^2\right]}\\
&=- \frac{4\pi^2 a^4  z_3 z_4 z''^{2}}{(4\pi^2)^4 r_{34}^{2}(1+\epsilon)^2} \frac{ \log \left(\frac{z''^2 \epsilon  (2+\epsilon)}{(1+\epsilon)^2 \left(x_3'''^2+z''^2\right)}\right)}{x_3'''^2 (\epsilon +1)^2+z''^2},\notag
\end{align}
which for small values of $\epsilon$ reduces to 
\begin{equation}
J_4^{d} = - \frac{4\pi^2 a^4  z_3 z_4 z''^{2}}{(4\pi^2)^4 r_{34}^{2}}~\frac{ \log \left(\frac{2 z''^2}{x_3'''^2+z''^2}\right)+\log (\epsilon )}{x_3'''^2+z''^2}+{\mathcal{O}}(\epsilon).
\end{equation}
The same steps done above, applied on $J_4^r$, lead to
\begin{align}
\begin{split}
J_4^r &= \frac{4\pi^2 a^4  z_3 z_4 z''^{2}}{(4\pi^2)^4 r_{34}^{2}} \int_{0}^{\infty}\mathrm{d}s_1\mathrm{d}s_2\mathrm{d}s_3~\frac{\delta(1- \sum^3_{i=1}s_i)}{\left[ (s_1+s_2)s_3 x_3'''^2 + (s_2 + s_3)s_1 z''^2 + s_2 z''^2\right]} \\
&=  \frac{4\pi^2 a^4  z_3 z_4 z''}{(4\pi^2)^4 r_{34}^{2}} \int_{0}^{1}\mathrm{d}s_3 \frac{\mathrm{arctanh} \sqrt\frac{(1-s_3)z''^2}{s_3 x_3'''^2 + z''^2}}{\sqrt{(1-s_3)(s_3 x_3'''^2 + z''^2)}}.
\end{split}
\end{align}
Then, the substitution 
\begin{equation}
t = \sqrt\frac{(1-s_3)z''^2}{s_3 x_3'''^2 + z''^2}    
\end{equation}
 yields a simpler form of the integral 
\begin{equation}
\label{J4rint}
J_4^r =  \frac{8\pi^2 a^4  z_3 z_4}{(4\pi^2)^4 r_{34}^{2}} \int_{0}^{1}\mathrm{d}t \frac{\mathrm{arctanh} t}{ \alpha^2 t^2 + 1},
\end{equation}
where we defined $\alpha = |x'''_3|/ z''$. This integral admits a closed-form solution in terms of the dilogarithm $\mathrm{Li}_2$:
\begin{equation}
J_4^r =  \frac{4\pi^2 a^4  z_3 z_4}{(4\pi^2)^4 r_{34}^{2}}~\frac{1}{\alpha}\mathrm{Im} \left\{ \mathrm{Li}_2 \left(\frac{\mathrm{i} \alpha -1}{\mathrm{i} \alpha +1}\right) \right\}\mkern-2mu.
\end{equation}
However, in order to compute the complete four-point function (\ref{K_4c}), the integral form (\ref{J4rint}) will be better suited. 
\paragraph{$\mathbf{\Delta=2}$.}
For $\Delta=2$, the integral (\ref{J4nu}) becomes instead
\begin{equation}\notag
J_4 = \frac{a^4 (16 z_3 z_4)^2 z''^{4}}{2 (4\pi^2)^4 r_{34}^{4}} \int_{-\infty}^{\infty} \mathrm{d}^3 y ~ \mathrm{d}w ~ \frac{w^4 [(x_3'''-y)^2 + w^2]^{-2}}{[(y^2 + (z''-w)^2 + \epsilon Q)(y^2 + (z''+w)^2+\epsilon Q)]^2}.
\end{equation}
As usual, let us introduce the Schwinger parameters:
\begin{align}
\begin{split}
J_4 = &\frac{a^4 (16 z_3 z_4)^2 z''^{4}}{2 (4\pi^2)^4 r_{34}^{4}}\int_{0}^{\infty}\mathrm{d}t_1\mathrm{d}t_2\mathrm{d}t_3 ~t_1 t_2 t_3 \int_{-\infty}^{\infty} \mathrm{d}^3 y ~ \mathrm{d}w ~w^4~ e^{-(t_1+t_2) \epsilon z''^2}\\
\times &~e^{-t_1 (z''-w)^2 -t_2 (z''+w)^2 - (\epsilon t_1 + \epsilon t_2 + t_3) w^2- (t_1+t_2)(1+\epsilon) y^2 - t_3 (x_3''' - y)^2}.
\end{split}
\end{align}
Integrating over the nonradial coordinates, and subsequently substituting $t_1 = s s_1,t_2 = s s_2, t_3=(1+\epsilon)s s_3$, yields
\begin{align}\notag
\begin{split}
J_4 =& \frac{a^4 \pi^{\frac32} (16 z_3 z_4)^2 z''^{4} (1+\epsilon)^{\frac32}}{2 (4\pi^2)^4 r_{34}^{4}}~\partial_\gamma^2|_{\gamma =0} \int_{0}^{\infty}\mathrm{d}s_1\mathrm{d}s_2\mathrm{d}s_3\mathrm{d}s ~s^{3/2} ~s_1 s_2 s_3 \int_{-\infty}^{\infty} \mathrm{d}w  ~\delta\left(1-\sum_{i=1}^3 s_i\right) \\
\times &~e^{- s(s_1+s_2)s_3(1+\epsilon) x_3'''^2-s (s_1 (z''-w)^2 +s_2 (z''+w)^2 + (\epsilon + s_3 + \gamma) w^2)-s (s_1+s_2) \epsilon z''^2}.
\end{split}
\end{align} 
Additionally, above we also introduced an auxiliary parameter $\gamma>0$ to get rid of the $w^4$ factor. Then, further evaluation of the integral leads to
\begin{align}\notag
\begin{split}
J_4=&  \frac{a^4 \pi^{2} (16 z_3 z_4)^2 z''^{4} (1+\epsilon)^{\frac32}}{2 (4\pi^2)^4 r_{34}^{4}}~\partial_\gamma^2|_{\gamma =0}\frac{1}{\sqrt{1+\gamma +\epsilon}}~ \int_{0}^{1}\mathrm{d}s_3\int_{0}^{1-s_3}\mathrm{d}s_1 ~s_1  s_3 (1-s_1-s_3) \\
\times &\left[(1-s_3)s_3(1+\epsilon) x_3'''^2+  \frac{4 s_1 (1-s_1-s_3) + (1-s_3) (\epsilon +s_3+ \gamma)}{1+\gamma+\epsilon}z''^2+ (1-s_3) \epsilon z''^2\right]^{-2}\mkern-5mu.
\end{split}
\end{align} 
Differentiating twice by $\gamma$, setting $\gamma=1$, and integrating over $s_1$, yields
\begin{equation}
J_4 =  \frac{16 a^4 \pi^{2} (z_3 z_4)^2 z''^{4} }{ (4\pi^2)^4 r_{34}^{4}} ~ \int_{0}^{1}\mathrm{d}s_3~\frac{ (1-s_3) s_3 (1+\epsilon)}{ \left(s_3 \left(x_3'''^2 (1+\epsilon)^2+z''^2\right)+z''^2 \epsilon  (2+ \epsilon)\right)^2}.
\end{equation}
Eventually, after integrating over $s_3$ and taking only the leading orders in $\epsilon$, one gets
\begin{equation}
J_4 = -\frac{16 a^4 \pi^{2} (z_3 z_4)^2 z''^{4} }{ (4\pi^2)^4 r_{34}^{4}}~\frac{ 2+ \log \left(\frac{2z''^2}{x_3'''^2+z''^2}\right)+ \log \epsilon  }{ \left(x_3'''^2+z''^2\right)^2} + \mathcal{O}(\epsilon).
\end{equation}
\paragraph{Recovering the full covariance.}
To get back the explicit covariant form, let us first note that
\begin{equation}
  \frac{4 z''^2}{x_3'''^2 + z''^2} = r_{34}^2 \bar{K}_{x x_3} \bar{K}_{x x_4} \equiv \frac{4}{\alpha^2 + 1},
\end{equation} 
which is manifestly covariant. Therefore, it follows that
\begin{equation}
J_4 = \frac{a^4  \pi^2}{(4 \pi^2)^4}   K^\Delta_{x x_3} K^\Delta_{x x_4}\left(\log(\alpha^2+1) -\log 2\epsilon + \Delta J_4(\alpha^2) \right)\mkern-3mu,
\end{equation}
where
\begin{equation}
\Delta J_4(\alpha^2)= \begin{cases} 2 (\alpha^2 + 1)\int_0^1 \mathrm{d}t~ \frac{\mathrm{arctanh}\, t}{\alpha^2 t^2 +1}&\mbox{for } \Delta =1, \\
 -2 &\mbox{for }\Delta=2. \end{cases}
\end{equation}

\paragraph{Attaching the missing legs.}
In order to obtain the complete one loop diagram $\mathcal{K}_4$ given in eq. (\ref{K_4c}), we still need to perform the remaining integral
\begin{equation}
\mathcal{K}_4 =  \int \mathrm{d}^4x \sqrt{|g(x)|} \Lambda (x_1,x,\Delta) \Lambda (x_2,x,\Delta) \mathcal{J}_4(x_3,x_4,x,\Delta).
\end{equation}
Let us send $z_1, z_2$ to the boundary, reducing $\mathcal{K}_4$ to
\begin{equation}
\mathcal{K}_4\sim K_4 = \frac{ a^4 (16 z_1 z_2 z_3 z_4)^\Delta \pi^2}{(4 \pi^2)^6}  \int \mathrm{d}^4x ~  \frac{ z^{4\Delta-4} \left(\log\frac{\alpha^2+1}{ 2\epsilon} + \Delta J_4(\alpha^2) \right)}{\Pi_{i=1}^4[(x_i-x)^2 + z^2]^\Delta}.
\end{equation}
In analogy to what was done above, let us translate $x_k,~k = 1,\dots,4$ by $(0,- x^i_4)$ (denoted by primes), invert all points (denoted by double primes), and then make the substitution $(z'',x^{i''})= (z, x^i + x^{i''}_3)$. Then, the above expression further simplifies to
\begin{equation}
\label{K4explicit}
K_4 =\frac{ a^4 (16 \Pi_{i=1}^4z_i)^\Delta \pi^2}{2 (4 \pi^2)^6 (x_1' x_2' x_3')^{2\Delta}}  \int_{-\infty}^{\infty} \frac{\mathrm{d}^3x ~\mathrm{d}z}{[x^2 + z^2]^\Delta}  \frac{ z^{4\Delta-4} \left[\log(\frac{x^2}{z^2}+1) -\log 2\epsilon + \Delta J_4(\frac{x^2}{z^2}) \right]}{[(x'''_1-x)^2 + z^2]^\Delta [(x'''_2-x)^2 + z^2]^\Delta },
\end{equation}
with $x^{i'''}_1 = x^{i''}_1 - x^{i''}_3$ and $x^{i'''}_2 = x^{i''}_2 - x^{i''}_3$. In order to solve $K_4$, let us first introduce a generating function with parameters $\gamma,t >0$:
\begin{equation}
\label{genexphi}
\phi_\Delta^{\gamma,t} =\frac{1}{2 x_1'^{2\Delta} x_2'^{2\Delta} x_3'^{2\Delta}}\int_{-\infty}^{\infty} \mathrm{d}^3x ~\mathrm{d}z~  \frac{ z^{4\Delta-4 + 2 \gamma}}{[(x'''_1-x)^2 + z^2]^\Delta [(x'''_2-x)^2 + z^2]^\Delta [t^2 x^2 + z^2]^{\Delta + \gamma}},
\end{equation}
comprising all different cases in eq. (\ref{K4explicit}). Indeed, the integral (\ref{K4explicit}) can be written as
\begin{equation}
\label{K4qfinal}
K_4 =\frac{ a^4 (16~ \Pi _{i=1}^4z_i)^\Delta \pi^2}{ (4 \pi^2)^6} \left[- \partial_\gamma \phi_\Delta^{\gamma,t=1} |_{\gamma=0} - \log\frac{\epsilon}{2} ~\phi_\Delta^{\gamma=0,t=1} + \Delta K_4 \right]\mkern-3mu, 
\end{equation} 
with
\begin{equation}
\Delta K_4= \begin{cases} 2 \int_0^1 \mathrm{d}t~\phi_{\Delta=1}^{\gamma=0,t} \mathrm{arctanh}\, t &\mbox{for } \Delta =1, \\
 -2 \phi_{\Delta=2}^{\gamma=0,t=1} &\mbox{for }\Delta=2. \end{cases}
\end{equation}
Introducing Schwinger parameters, integrating over spatial coordinates and making the usual substitution $t_i = s s_i$, yields
\begin{equation}\notag
\phi_\Delta^{\gamma,t} =\frac{\pi^{3/2}\Gamma(\Delta)^{-1}\Gamma(2\Delta + \gamma - \frac{3}{2}) }{2 \Gamma(\Delta+\gamma) (\eta\zeta ~\Pi_{i<j} r_{ij})^{\frac{2}{3}\Delta}}\int_{0}^{\infty} \frac{\left(\Pi_{i=1}^3\mathrm{d}s_i\right)}{(s_1 s_2 s_3)^{1-\Delta}}~ \frac{  s_3^{\gamma}(\sum_{i=1}^3 s_i)^{\frac{3}{2} - 2\Delta - \gamma}~\delta(1 - \sum_{i=1}^3 s_i)}{(1 + s_3(t^2-1))^{\frac{3}{2}-\Delta} [\frac{s_1 s_2 }{ \eta^2} + t^2 s_3(\frac{ s_1}{\zeta^2} + s_2)]^{\Delta}},
\end{equation}
where we reintroduced the conformal invariants already defined in eq. (\ref{confinvariantlist}). Making another substitution $s_1\rightarrow s s_1,s_2\rightarrow s s_2,s_3 \rightarrow s$ and integrating over $s$ further simplifies the integral to
\begin{equation}
\label{gammatintegral}
\phi_\Delta^{\gamma,t} =\frac{\pi^{3/2}\Gamma(\Delta)^{-1}\Gamma(2\Delta + \gamma - \frac{3}{2}) }{2 \Gamma(\Delta+\gamma) (\eta\zeta ~\Pi_{i<j} r_{ij})^{\frac{2}{3}\Delta}}\int_{0}^{\infty} \mathrm{d}s_1 \mathrm{d}s_2 ~ \frac{ (s_1 s_2)^{\Delta-1} (1+s_1+s_2)^{\frac{3}{2}-2\Delta-\gamma }}{(t^2 + s_1+s_2)^{\frac{3}{2}-\Delta}  [\frac{s_1 s_2 }{ \eta^2} +t^2(\frac{ s_1}{\zeta^2} + s_2)]^{\Delta}}.
\end{equation}
The change of variables $s_1 =t^2 s r, s_2 =t^2 s (1-r)$ compactifies one integration region, which leads to
\begin{equation}
\label{gammatintegral2}
\phi_\Delta^{\gamma,t} =\frac{\pi^{3/2}\Gamma(\Delta)^{-1}\Gamma(2\Delta + \gamma - \frac{3}{2}) }{2 \Gamma(\Delta+\gamma) (\eta\zeta ~\Pi_{i<j} r_{ij})^{\frac{2}{3}\Delta}}\int_{0}^{\infty} \mathrm{d}s \int_0^1 \mathrm{d}r ~ \frac{ (s r (1-r))^{\Delta-1} (1+ t^2 s)^{\frac{3}{2}-2\Delta-\gamma }}{t^{3-2\Delta} (1 + s)^{\frac{3}{2}-\Delta}  [s \frac{r(1-r) }{ \eta^2} +\frac{ r}{\zeta^2} + 1- r]^{\Delta} }.
\end{equation}
Note that, as one might expect, eq. (\ref{gammatintegral2}) is invariant under the exchanges $x_1 \leftrightarrow x_2$ and $x_3 \leftrightarrow x_4$. For example, $x_1 \leftrightarrow x_2$ yields $\eta \rightarrow \frac{\eta}{\zeta}, \zeta \rightarrow \frac{1}{\zeta}$, and then invariance of the above formula follows after a change of variables in $r$.
\paragraph{Term-by-term computation.}
Setting $\gamma=0,t=1$ in eq. (\ref{gammatintegral}) should lead to the four-point function at tree level, which was computed earlier. Indeed, integrating over $s_1$ yields
\begin{align}
\begin{split}
\phi_\Delta^{\gamma=0,t=1} =&\frac{ \pi^{3/2} \Gamma(2\Delta  - \frac{3}{2}) ~\eta^{2\Delta}\zeta^{2\Delta}}{ 2 \Gamma(2\Delta) (\eta\zeta ~\Pi_{i<j} r_{ij})^{\frac{2}{3}\Delta}}\int_{0}^{\infty} \frac{\mathrm{d}s_2}{s_2^{1-\Delta}} ~ \frac{  (1+s_2)^{-\Delta}}{(\eta^2 s_2+ \zeta^2)^\Delta}\\
&\times  ~_2 F_1 \left[\Delta,\Delta;2\Delta; 1-\frac{s_2 \eta ^2 \zeta^2 }{ (1+s_2) \left(\eta ^2 s_2+\zeta^2\right)}\right]\mkern-2mu,
\end{split}
\end{align}
corresponding to $I_4$ given in eq. (\ref{finalformula}) up to a Pfaff transformation of the hypergeometric function. This result, together with eq. (\ref{K4qfinal}), confirms the expectation that the UV divergence can be completely absorbed in the coupling constant $\lambda$. The first term in eq. (\ref{K4explicit}) is given by
\begin{equation}
- \partial_\gamma \phi^{\gamma,t=1}_\Delta |_{\gamma= 0}= I_4 \left( \psi(\Delta) - \psi(2\Delta-\frac{3}{2})\right) +\frac{\pi^{\frac{3}{2}}\Gamma(2\Delta - \frac{3}{2}) }{2\Gamma(\Delta)^2 } L_4 (\eta,\zeta),
\end{equation}
  where $\psi(x)$ denotes the digamma function and where
\begin{equation}
\label{Lfour}
L_4 (\eta,\zeta)=\frac{1}{(\eta\zeta ~\Pi_{i<j} r_{ij})^{\frac{2}{3}\Delta}}\int_{0}^{\infty} \mathrm{d}s \int_0^1 \mathrm{d}r ~  \frac{ (s r (1-r))^{\Delta-1}\log (1+s)}{ (1+s)^{\Delta} [\frac{s r (1-r) }{ \eta^2} +\frac{r}{\zeta^2} +1- r]^{\Delta}}.
\end{equation}  
In the $\Delta=1$ case, eq. (\ref{K4explicit}) contains also the term

\begin{equation}
 2  \int_{0}^{\infty} \mathrm{d}t~ \mathrm{arctanh}~t ~\phi_{\Delta=1}^{\gamma=0,t} = \pi^2 L'_4 (\eta,\zeta),
\end{equation}
where
\begin{equation}
\label{L4primefirstdef}
L'_4 (\eta,\zeta) = \frac{1 }{ (\eta\zeta ~\Pi_{i<j} r_{ij})^{\frac{2}{3}}}\int_0^1 \mathrm{d}t \int_{0}^{\infty} \mathrm{d}s \int_0^1 \mathrm{d}r ~  \frac{ \mathrm{arctanh}~ t}{t \sqrt{(1 + s) (1+t^2 s)} [\frac{s r (1-r) }{ \eta^2} +\frac{ r}{\zeta^2} + 1-r]}.
\end{equation}
By means of the relation
\begin{equation}
    \frac{\partial}{\partial s} \int_0^1 \mathrm{d}t~ \frac{\mathrm{arctanh}~t}{t \sqrt{1 + t^2 s}} = \frac{\partial}{\partial s} \int_1^\infty \mathrm{d}\lambda~ \frac{\log (1+ \lambda s)}{4 \lambda \sqrt{1 + \lambda s}} = - \frac{1}{4} \frac{\log (1+s)}{s \sqrt{1+s}},
\end{equation}
which holds for all $s \geq 0$, one can rewrite $L'_4(\eta,\zeta)$ as
\begin{equation}
\label{Lfourprime}
L'_4 (\eta,\zeta) = \frac{1 }{ (\eta\zeta ~\Pi_{i<j} r_{ij})^{\frac{2}{3}}}\int_1^\infty \mathrm{d}\lambda \int_{0}^{\infty} \mathrm{d}s \int_0^1 \mathrm{d}r ~  \frac{ \log(1+ \lambda s) }{4 \lambda \sqrt{(1 + s) (1+\lambda s)} [\frac{s r (1-r) }{ \eta^2} +\frac{ r}{\zeta^2} + 1-r]}.
\end{equation}
Putting everything together, we find that (cf. eq. \eqref{K4qfinal})
\begin{equation}
\label{K4final}
K_4 = \frac{a^4 4^{2\Delta}(\Pi_{i=1}^4 z_i)^\Delta \pi^2}{(4\pi^2)^6} \left[  \left(\psi(\Delta) - \psi(2\Delta- \frac{3}{2}) - \log \frac{\epsilon}{2}  \right) I_4 + \pi^{\frac{3}{2}} \frac{\Gamma (2\Delta - \frac{3}{2})}{2\Gamma(\Delta)^2} L_4 + \Delta K_4\right]
\end{equation}
with
\begin{equation}
\Delta K_4= \begin{cases} \pi^2 L'_4(\eta,\zeta) &\mbox{for } \Delta =1, \\
 -2 I_4 &\mbox{for }\Delta=2. \end{cases}
\end{equation}
The quantities $L_4(\eta,\zeta)$ and $L'_4(\eta,\zeta)$ are given respectively in eq. (\ref{Lfour}) and eqs. (\ref{L4primefirstdef},\ref{Lfourprime}). The above result, together with the loop-corrected two-point functions given in section \ref{2ptfnchapter}, are the key new results of this paper. Eq. (\ref{K4final}) contains the complete s-channel contribution to the one-loop four-point function in AdS for $\Delta=1$ as well as $\Delta=2$. The latter was briefly reported in ref. \cite{Bertan:2018khc}. The t-channel can be simply recovered by an exchange $x_1 \leftrightarrow x_4$, which turns out to be equivalent to $\eta \leftrightarrow \zeta$. Analogously, the u-channel corresponds to the exchange $x_2 \leftrightarrow x_4$, which, in terms of $\eta$ and $\zeta$, translates to $\eta \rightarrow \frac{1}{\eta}$, $\zeta \rightarrow \frac{\zeta}{\eta}$. In the next sections we will relate these results to the conformal block expansion which, in turn, defines the dual conformal field theory. This will be done with the help of a  short-distance expansion of eq. (\ref{K4final}).

\section{Conformal blocks and anomalous dimensions}\label{chapope}
In this section we would like to compare the results of the explicit AdS computations with the general expectations arising from the operator product expansion. First of all, let use summarize the final result of the bulk computation. 
The Witten two-point function is given by\footnote{The remaining diagrams including tadpoles and sunset diagrams are proportional to the mass shift and thus merely contribute to the relation between renormalized and bare mass of the bulk scalar, as explained in section \ref{ssd}.}
\begin{equation}
    \langle \bar{\phi}(x_1) \bar{\phi}(x_2) \rangle =\diagramoneb =  \frac{N_\phi}{r_{12}^{2\Delta}},
\end{equation}
where $N_\phi = \frac{a^2 (2 \Pi_{i=1}^2 z_i)^\Delta}{ 4\pi^2}$. To continue we take $z_i\equiv z$ for all external legs. 
Then, for the values $\Delta=1,2$, the Witten four-point function reads (cf.  eq. (\ref{4ptdiagrams}))
\begin{align}
    \begin{split}
\label{4ptftofirstorder}
    \langle \bar{\phi}(x_1)& \bar{\phi}(x_2) \bar{\phi}(x_3) \bar{\phi}(x_4) \rangle = \diagramtwob  \\[0.5em]
    &+\lambda~ \diagramtwoc + \mathcal{O}\left( \lambda^3 \right)\\[0.5em]
    &=\frac{N_\phi^2}{(r_{12}r_{34})^{2\Delta}} \left[1+  v^\Delta  + \frac{v^\Delta}{(1-Y)^\Delta} +\lambda\frac{v^\Delta(2\Delta - 1)}{4\pi^2 }  I_0 \right. \\[0.5em]
    \; &~~~~~~~~~~~~~~~~~~\left. +\lambda^2 \frac{3\psi_0}{2} \frac{ v^\Delta(2\Delta - 1)}{4^{3}\pi^4 } I_0+ \lambda^2\frac{v^\Delta(2\Delta - 1)}{4^{4}\pi^{4} }K_0 \right]+ \mathcal{O}\left( \lambda^3 \right)\mkern-3mu,
        \end{split}
\end{align}
where 
\begin{align}
\label{def4ptf}
    \begin{split}
\psi_0 &= \psi(\Delta) - \psi(2\Delta- \frac{3}{2}) - \log \frac{\epsilon}{2}, \\
    I_0 &=  \sum_{n,m=0}^\infty \frac{ Y^m v^n}{ m! (n!)^2} \frac{\Gamma(n+\Delta)^2 \Gamma(n+m+\Delta)^2}{\Gamma(2n + m + 2\Delta)} \psi_{nm},\\
     \psi_{nm} &= \psi(n+1) -\frac12 \log v  - \psi(n+\Delta) - \psi(n+\Delta + m) + \psi(2n + 2\Delta +m),
\end{split}
\end{align}
and
\begin{align}
\label{defK0}
K_0= - 12 (\Delta-1) I_0 + \sum_{\substack{\pi(x,y,z)\\ x\rightarrow v,y\rightarrow 1-Y,z\rightarrow 1}} \begin{cases} L_0(x,y,z) + 2 L'_0(x,y,z) &\mbox{for } \Delta =1, \\
 L_0(x,y,z) &\mbox{for }\Delta=2. \end{cases}
\end{align}
The above sum runs over the three cyclic permutations $\pi$ of the variables $x,y,z$, namely over the s-, u- and t-channel. The appearing quantities are defined as
\begin{equation}
\label{L0}
L_0(x,y,z) =\int_{0}^{\infty} \mathrm{d}s \int_0^1 \mathrm{d}r ~  \frac{ (s r (1-r))^{\Delta-1}\log (1+s)}{ (1+s)^{\Delta} [s r (1-r) x + r y  + (1- r)z]^{\Delta}},
\end{equation} 
and
\begin{equation}
\label{L0prime1}
L'_0(x,y,z) =  \int_0^1 \mathrm{d}t \int_{0}^{\infty} \mathrm{d}s \int_0^1 \mathrm{d}r ~  \frac{ \mathrm{arctanh}~ t}{t \sqrt{(1 + s) (1+t^2 s)} [s r (1-r) x + r y + (1-r)z]},
\end{equation}
or, equivalently,
\begin{equation}
\label{L0prime2}
L'_0(x,y,z) =  \int_1^\infty \mathrm{d}\lambda \int_{0}^{\infty} \mathrm{d}s \int_0^1 \mathrm{d}r ~  \frac{ \log(1+ \lambda s) }{4 \lambda \sqrt{(1 + s) (1+\lambda s)} [s r (1-r) x +r y + (1-r)z]}.
\end{equation}
Note that, in this form, the Witten four-point function (\ref{4ptftofirstorder}) explicitly shows covariance under conformal symmetry.

The logarithmically divergent terms in eq. (\ref{4ptftofirstorder}) can be absorbed in the coupling constant. Indeed, the renormalized coupling constant obtained through a nonminimal subtraction is given by
\begin{equation}
    \lambda = \lambda_R - \frac{3 \lambda_R^2}{32 \pi^2 } \psi_0+ \mathcal{O}(\lambda_R^3).
\end{equation}
Varying the coupling constant with respect to the square root of $\epsilon$ leads to the beta function
\begin{equation}
    \beta(\lambda) \equiv \sqrt{\epsilon} \frac{\partial \lambda}{\partial \sqrt{\epsilon}} =  \frac{3\lambda^2}{16\pi^2 } + \mathcal{O}(\lambda^3) 
\end{equation}
known from standard QFT literature. In what follows, we will use the renormalized coupling constant $\lambda_R$. In addition, we will fine-tune the renormalized mass such that the renormalized bulk theory is conformally coupled to geometry. Since the mass has a nonvanishing bulk beta function this is not an bulk RG-invariant statement. One might thus worry that such a theory cannot be dual to a conformal theory on the boundary. This apparent contradiction is, however, resolved by noting that three-dimensional scale transformations on the boundary correspond to an isometry in the four-dimensional bulk. 

\subsection{Operator product expansion}
Our final goal is to compare our results on the AdS side, that is, correlations functions evaluated on the boundary, with the double OPE of the full four-point function on the boundary itself. In order to do so, we have to consider the separate limits $v,Y \rightarrow 0$ of the whole holographic four-point function (\ref{4ptftofirstorder}). For the cross diagram $I_0$, the required expansion is already given to all orders in eq. (\ref{def4ptf}). On the other hand, for $K_0$ given in eq. (\ref{defK0}), the computation is more elaborate. The adopted procedure is described in detail in appendix \ref{app:limits}.

The bulk scalar is dual to a real scalar operator $\mathcal{O}_\Delta$ of weight $\Delta$. In the leading approximation, $\mathcal{O}_\Delta$ is a generalized free field and the OPE of $\mathcal{O}_\Delta$ with itself contains double-trace operators $\mathcal{O}_{n,l}$ of all even spins $l$ with conformal dimension $2\Delta+2n+l$. Schematically, these operators are of the form $\mathcal{O}_{n,l}=:\mkern-4mu\mathcal{O}_\Delta \square^n \pl^l \mathcal{O}_\Delta\mkern-4mu:$. The OPE reads
\begin{align}
    \mathcal{O}_\Delta \mathcal{O}_\Delta &=1+\sum_{n,l} \mathcal{A}^{1/2}_{n,l}\, \mathcal{O}_{n,l}\,,
\end{align}
where $\mathcal{A}^{1/2}_{n,l}$ are the OPE coefficients. 
The cubic vertex is absent in our model, and for this reason $\mathcal{O}_\Delta$ itself does not show up in the OPE. In the leading approximation, the four-point function is given by the disconnected contributions, coming from the product of two-point functions
\begin{align}
    \langle \mathcal{O}_\Delta \mathcal{O}_\Delta \mathcal{O}_\Delta \mathcal{O}_\Delta\rangle&= \frac{1}{(r_{12}r_{34})^{2\Delta}}\left(1+ v^\Delta +\frac{v^\Delta}{(1-Y)^\Delta}\right)\mkern-3mu,
\end{align}
where we drop $N_{\phi}^2$ from eq. (\ref{4ptftofirstorder}), being an overall factor. 
On the other hand, the four-point function has a conformal block expansion\footnote{We use the recursion relations from ref. \cite{Dolan:2000ut}. The conformal block $G_{\Delta,l}$ of the spin-$l$ operator with weight $\Delta$ begins with $v^{(\Delta-l)/2}(Y^l2^{-l}+...)$. See also appendix \ref{app:OPE}.} 
\begin{align}
    \langle \mathcal{O}_\Delta \mathcal{O}_\Delta \mathcal{O}_\Delta \mathcal{O}_\Delta\rangle&=\frac{1}{(r_{12}r_{34})^{2\Delta}}\left(G_{0,0}+\sum_{n,l} \mathcal{A}_{n,l} G_{n,l} \right)\mkern-3mu,
\end{align}
where $G_{0,0}$ is the contribution of the unit operator and the squares of the OPE coefficients are given in appendix \ref{app:OPE}  for any conformal weight $\Delta$ and any dimension $d$, see also ref. \cite{Fitzpatrick:2011dm}. Below we consider separately the two cases of interest for us.

The connected bulk diagrams result in corrections to the OPE data: the OPE coefficients and anomalous dimensions, which we would like to extract. It is useful to consider the squares of the OPE coefficients $\mathcal{A}_{n,l}$ as functions of the conformal dimension:
\begin{align}\label{Agamma}
    \mathcal{A}_{\Delta_{n,l} + \gamma^{(1)}_{n,l}+ \gamma^{(2)}_{n,l}+...}&=A_{n,l}+\gamma^{(1)}_{n,l}A^{(1)}_{n,l}+ \left(\gamma^{(2)}_{n,l}A^{(1)}_{n,l}+\frac12 (\gamma^{(1)}_{n,l})^2A^{(2)}_{n,l}\right)+ \dots \,,
\end{align}
where it is assumed that $\gamma^{(k)}_{n,l}$ is of order $\lambda_R^k$, while $A_{n,l},A^{(k)}_{n,l}$ are just numbers. 
Therefore, the conformal block expansion, up to second order in the coupling constant, reads
\begin{align}\notag
    \begin{aligned}
    \langle \mathcal{O}_\Delta & \mathcal{O}_\Delta \mathcal{O}_\Delta \mathcal{O}_\Delta\rangle=G_{0,0}+\sum_{n,l} A_{n,l} G_{n,l} + \sum_{n,l}\gamma^{(1)}_{n,l}\left(A^{(1)}_{n,l}G_{n,l}+A_{n,l}G_{n,l}'\right) \\
    +&\sum_{n,l}\left[ \gamma^{(2)}_{n,l}(A_{n,l}G_{n,l}'+A_{n,l}^{(1)} G_{n,l})+\frac12 (\gamma^{(1)}_{n,l})^2\left(A^{(2)}_{n,l}G_{n,l}+A_{n,l}G_{n,l}''+2A^{(1)}_{n,l}G_{n,l}'\right)\right]+\mathcal{O}(\lambda_R^3),
    \end{aligned}
\end{align}
where $G_{n,l}'$ and $G_{n,l}''$ are the derivatives with respect to the conformal dimension evaluated at the free-field value of the double-trace operator's conformal dimension, i.e., at $2\Delta+2n+l$. In what follows we perform the conformal block expansion of the bulk results and extract the OPE data. We first discuss the $\Delta=2$ case and only afterwards the $\Delta=1$ case, since the former leads to simpler results.

\paragraph{$\mathbf{\Delta=2}$.}
The result of the bulk computation is obtained from eq. (\ref{4ptftofirstorder}) by setting $\Delta=2$. The zeroth-order OPE coefficients correspond to the disconnected part and follow from the general result discussed above
\begin{align*}
   A_{n,l}&=\frac{2^{-l-4 n} \Gamma \left(l+\frac{3}{2}\right) \Gamma \left(n+\frac{3}{2}\right) \Gamma (n+2) \Gamma (l+n+2) \Gamma \left(l+n+\frac{5}{2}\right) \Gamma (l+2 n+3)}{\Gamma (l+1) \Gamma (n+1)^2 \Gamma \left(l+n+\frac{3}{2}\right)^2 \Gamma \left(l+2 n+\frac{5}{2}\right)}.
\end{align*}
The first order anomalous dimensions are easy to extract from $I_0$ in eq. (\ref{4ptftofirstorder}):
\begin{align}
    \gamma := \gamma_{n,l=0}^{(1)}&=-\frac{\lambda_R}{16 \pi ^2}, &&\gamma_{n,l>0}^{(1)}=0.
\end{align}
Only the scalar operators $:\mkern-6mu\mathcal{O}_\Delta \square^n \mathcal{O}_\Delta\mkern-6mu:$ receive anomalous dimensions and, for the simplest quartic interaction as considered here, the anomalous dimension does not depend on $n$, see also ref. \cite{Heemskerk:2009pn}. It is known that such an interaction does not induce anomalous dimensions for the operators with $l>0$. The OPE coefficients are not so illuminating, but we find them to be in accordance with refs. \cite{Heemskerk:2009pn,Fitzpatrick:2011dm} (note that $\gamma^{(1)}_{n,l}$ does not depend on $n$ and, according to eq. \eqref{Agamma}, is factored out of the OPE coefficients):
\begin{align}\label{Aonefree}
    A^{(1)}_{n,l}&= \frac12 \frac{\pl}{\pl n} A_{n,l}\,,&& l=0\,.
\end{align}
Additionally, for operators with spin, the OPE coefficients can be determined only at the second order since $\gamma^{(1)}_{n,l>0}=0$, as is clear from eq. (\ref{Agamma}). 

At second order the first few anomalous dimensions read (see appendix \ref{app:OPE} for a more detailed table) 
\begin{align*}
    \gamma^{(2)}_{0,0}&=\frac53\gamma^2, &&\gamma^{(2)}_{0,2}=-\frac{1}{20}\gamma^2, &&\gamma^{(2)}_{0,4}=-\frac{1}{140}\gamma^2\,, && \gamma^{(2)}_{0,6}=-\frac{1}{504}\gamma^2\,,\\
    \gamma^{(2)}_{1,0}&=\frac{46}{15}\gamma^2\,, && \gamma^{(2)}_{1,2}=-\frac{107}{1260}\gamma^2\,, &&\gamma^{(2)}_{1,4}=-\frac{19}{1260}\gamma^2\,,\\
    \gamma^{(2)}_{2,0}&=\frac{113}{28}\gamma^2\,, && \gamma^{(2)}_{2,2}=-\frac{269}{2520}\gamma^2\,. &&
\end{align*}
Note that the loop correction results in nonvanishing anomalous dimensions for spinning operators as well. Indeed, since there is no operator which saturates the unitarity bound, in our model no local stress tensor and not even a conserved current appears. Therefore, none of the operators is expected to be protected.

While most of the anomalous dimensions are quite complicated, by comparison with the expansion of eq. (\ref{4ptftofirstorder}) we found a simple formula for the leading-twist operators, i.e., for $:\mkern-4mu\mathcal{O}_\Delta \pl^l \mathcal{O}_\Delta\mkern-4mu:$ (being in the first Regge trajectory):
\begin{align}
\label{resultgamma}
    \gamma_{0,l}^{(2)}&=\gamma^2
    \begin{cases}
        \frac53 & \mathrm{for~}l=0,\\
        -\frac{6}{ (l+3) (l+2)(l+1) l} & \mathrm{for~}l>0.
    \end{cases}
\end{align}
The general pattern is that all operators with nonzero spin have negative anomalous dimensions, which should correspond to binding energies in the AdS dual picture. Only the scalar operators have positive anomalous dimensions, which is, however, a second-order effect as compared to  $\gamma^{(1)}_{n,l=0}$.

From eq. (\ref{resultgamma}) we can easily work out the conformal spin expansion of the anomalous dimensions of the operators belonging to the leading trajectory and notice the latter to be consistent with the general expectations \cite{Basso:2006nk,Alday:2015eya,Alday:2015ewa}
\begin{align}
    \gamma^{(2)}_{0,l>0}=\gamma^2\frac{3}{J^2 (1-J^2/2)}, && J^2=(l+\tau/2)(l+\tau/2-1)=(l+2)(l+1),
\end{align}
where the twist $\tau$ corresponds to $4$.

\paragraph{$\mathbf{\Delta=1}$.}
Here we set $\Delta=1$ in the result of the bulk computation \eqref{4ptftofirstorder}. The zeroth-order OPE coefficients are then given by\footnote{Note that there is small subtlety in taking $\Delta=1$ in the general formula \eqref{genAformula}, and the $n=0$ case is special.}
\begin{align*}
   A_{n>0,l}&=\frac{4 \Gamma\left(l+\frac{3}{2}\right) \Gamma\left(n+\frac{1}{2}\right) \Gamma(l+n+1) \Gamma(l+2 n+1)}{2^{l+4 n}\Gamma(l+1) \Gamma(n+1) \Gamma\left(l+n+\frac{3}{2}\right) \Gamma\left(l+2 n+\frac{1}{2}\right)}\,,& A_{n=0,l}&=\frac{2\sqrt{\pi } \Gamma(l+1)}{2^l\Gamma\left(l+\frac{1}{2}\right)}\,.
\end{align*}
The first-order anomalous dimensions are found to be
\begin{align}
    \gamma_{n=0,l=0}^{(1)}&=2\gamma, & \gamma_{n>0,l=0}^{(1)}&=\gamma, &&\gamma_{n,l>0}^{(1)}=0.
\end{align}
We observe a similar pattern as for $\Delta=2$, except that the anomalous dimension of the very first operator in the OPE, $:\mkern-4mu\mathcal{O}_\Delta^2\mkern-6mu:$, jumps from $\gamma$ to $2\gamma$. The first order OPE coefficients follow \eqref{Aonefree}, as expected.

At second order our results are more limited as compared to the $\Delta=2$ case, the reason being that we did not find an efficient expansion for the integrals $L_0'(x,y,z)$ in eq. (\ref{L0prime1}) at high orders in $v$ and $Y$. Nevertheless, the anomalous dimension of the operators on the first Regge trajectory can be determined to all orders in the spin
\begin{align}\label{anomalB}
    \gamma^{(2)}_{n=0,l}&=  \gamma^2\frac{-4 }{2 l+1}\psi^{(1)}(l+1)+ \gamma^2\begin{cases}
        -4 & \mathrm{for~} l=0,\\
        -\frac{2}{l(l+1) }, &\mathrm{for~} l>0,
    \end{cases}
\end{align}
where $\psi^{(1)}$ is the digamma function, which can be rewritten also as
\begin{align} 
    \frac{-4 }{2 l+1}\psi^{(1)}(l+1)=\frac{1}{2l+1}\left(-\frac{2\pi^2}{3 }+{4 H_l^{(2)}}\right)\mkern-3mu,
\end{align}
where $H_l^{(2)}=\sum_{k=1}^{l}k^{-2}$ are the generalized harmonic numbers. It is actually in this latter form that the anomalous dimensions emerge from the OPE expansion of the bulk integrals. The last term in eq. (\ref{anomalB}) results from all channels of $L_0(x,y,z)$ and the $s$-channel of $L_0'(x,y,z)$. In the large spin limit the anomalous dimension behaves as
\begin{align}
   \gamma^{(2)}_{n=0,l}/\gamma^2&= -4/l^2 + 4/l^3 - 10/(3 l^4)+\mathcal{O}(1/l^5).
\end{align}
It can also be seen that the anomalous dimension admits an expansion in terms of the conformal spin. We expect to find a series of the form \cite{Basso:2006nk,Alday:2015eya,Alday:2015ewa}
\begin{align}
    \gamma^{(2)}_{0,l}=\gamma^2\sum_{k=1} \frac{Q_k}{J^{2k}}, && J^2=(l+\tau/2)(l+\tau/2-1)=l(l+1),
\end{align}
where the twist $\tau$ is $2$ and $Q_k$ are coefficients to be determined. The last term in eq. (\ref{anomalB}) contributes with $-2$ to $Q_1$. It is interesting that the first term can also be expanded and the coefficients are related to the Euler-Ramanujan's harmonic number expansion into negative powers of the triangular numbers
\begin{align}
    Q_k&=(-)^{k+1} 2^{1-2 k} \left(\sum _{j=1}^k (-4)^j \binom{k}{j} B_{2 j}(\tfrac12) +1\right)\mkern-3mu,
\end{align}
where $B_{2j}(x)$ are the Bernoulli polynomials. 

The anomalous dimensions of the operators belonging to the subleading Regge trajectories do not have any  $\pi^2$ (or polygamma) contributions  and are listed in appendix \ref{app:OPE}. All of them are negative for $l>0$ and positive for $l=0$.

\newpage
\section{Conclusions}\label{conc}
In this paper we gave an analytic derivation of the two-loop correction to bulk- and boundary two-point functions for a conformally coupled interacting scalar field in Euclidean AdS, as well as the one-loop correction for the four-point boundary-to-boundary correlation function, by generalizing the usual flat space Feynman perturbation theory to AdS. The final result can be reduced to a single integral expression which is not given by elementary functions. The remaining integral can either be evaluated numerically or, more importantly, be evaluated analytically in a short-distance expansion on the boundary. We have then shown that the corresponding expansion coefficients fix uniquely the data for the conformal block expansion of the dual conformal field theory on the boundary of AdS and no contradiction arises despite subtleties with UV and (sometimes) IR divergencies. The structure of the dual CFT turns out to be that of a deformed generalized free field of dimension $\Delta=1$ and $\Delta=2$. The OPE of the CFT contains an infinite number of further primary double-trace operators which have anomalous dimensions and anomalous OPE coefficients that we are able to compute from our boundary correlation functions. This is the AdS equivalent of determining the masses and branching ratios in flat space-time. In order for the interpretation of our result to work out correctly in terms of a dual CFT, our loop corrected boundary correlations function have to pass some nontrivial consistency tests. For example, the first order anomalous dimension enters not just at tree-level, but also in the bulk four-point function at one loop multiplying $\log(v)^2$. Similarly, the conformal spin expansion \cite{Basso:2006nk,Alday:2015eya,Alday:2015ewa} implies a certain asymptotic fall-off behavior of the anomalous dimensions for large spin. All of these conditions are fulfilled by our bulk correlators. In addition, our bulk calculation gives manifestly finite results for all anomalous dimensions in terms of the renormalized bulk coupling, something that is more difficult to achieve in an approach that reconstructs the bulk correlators from the boundary CFT (e.g., refs. \cite{Aharony:2016dwx, Alday:2017xua,Aprile:2017bgs}).

\section*{Acknowledgments}
I.B. would like to thank Tom\'{a}\v{s} Proch\'{a}zka for assistance with Mathematica and Sebastian Konopka for discussions. I.S. would like to thank Stefan Theisen for discussions, Eric Perlmutter for helpful correspondence, and the AEI Potsdam for hospitality during part of this project. We are indebted to Hynek Paul for bringing to our attention the conjecture of ref. \cite{Heemskerk:2009pn}. This work was supported by the DFG Transregional Collaborative Research Centre TRR 33 and the DFG cluster of excellence ``Origin and Structure of the Universe". The work of E.S. was supported by the Russian Science Foundation grant 18-72-10123 in association with the Lebedev Physical Institute.

\begin{appendix}\label{app}
\renewcommand{\thesection}{\Alph{section}}
\renewcommand{\theequation}{\Alph{section}.\arabic{equation}}
\setcounter{equation}{0}\setcounter{section}{0}
\newpage
\section{Expansions in the conformal invariants}
\label{app:limits}
\setcounter{equation}{0}
In this appendix we explain how to evaluate the expansion of the integrals $L_0$ and $L_0'$ given in eqs. (\ref{L0}-\ref{L0prime2}) in powers of $v$ and $Y$, that is\footnote{To do so one has to substitute the variables $x,y,z$ with the conformal invariants $v,1-Y$ and $1$ for each channel in eqs. (\ref{L0}-\ref{L0prime2}). In order to simplify the notation we will simply take the subscript in $L_0$ and $L_0^{\prime}$ as a placeholder for the different channels $s,t,u$.}
\begin{equation}
    L_0 = \sum_{m,n=0}^{\infty}L_0^{(n,m)}(\log v)~ v^n Y^m,
\end{equation}
and analogously for $L_0^{\prime}$. Here, the coefficient functions $L_0^{(n,m)}(\log v)$ and $L_0^{\prime(n,m)}(\log v)$ are to be determined. It is possible to obtain these coefficient functions analytically up to reasonably high order with the help of Mathematica. For $L_0$, the implementation of the code is quite straightforward and works efficiently to high orders. For $L_0'$, however, this turns out to be a much more difficult task. Nevertheless, we able to provide a code which works up to a sufficient order for our purposes. In this setting, the upper bound of computable orders is set by  the t- and u-channel of $L_0'$, as will be explained later. 

\paragraph{The integral $\mathbf{L_0}$.}
We first discuss the simpler integral $L_0$ given by
\begin{equation}
\label{L0xyz}
L_0(x,y,z) =\int_{0}^{\infty} \mathrm{d}s \int_0^1 \mathrm{d}r ~  \frac{ (s r (1-r))^{\Delta-1}\log (1+s)}{ (1+s)^{\Delta} [s r (1-r) x + r y  + (1- r)z]^{\Delta}}.
\end{equation}

\noindent{\it The s-channel.}
The s-channel is given by $L_0(v,1-Y,1)$. Let us denote the associated integrand by $l_s(s,r;v,Y)$. Then, for each of the cases $\Delta=1,2$, the steps to follow are:
\begin{enumerate}
    \item Expansion in $Y$: $\qquad ~~~l_s(s,r;v,Y)= \sum_{m=0}^\infty l_s^{(m)}(s,r;v) ~Y^m$
    \item Integration over $s$: $\qquad l_s^{(m)}(r;v) = \int_0^\infty \mathrm{d}s ~l_s^{(m)}(s,r;v)$
    \item Expansion in $v$: $\qquad ~~~\;l_s^{(m)}(r;v) = \sum_{n=0}^\infty l_s^{(n,m)}(r;\log v)~ v^n$
    \item Integration over $r$: $\qquad  L_s^{(n,m)}(\log v) = \int_0^1 \mathrm{d}r~l_s^{(n,m)}(r;\log v)$
\end{enumerate}

\noindent{\it The t- and u-channels.}
For both the t- and u-channels of $L_0$, although the integrals being different, the employed procedure is the same. The integrals are respectively given by $L_0(1-Y,1,v)$ and $L_0(1,v,1-Y)$. Here, the steps to follow are similar as above, but with the order  of integration interchanged:
\begin{enumerate}
    \item Expansion in $Y$: $\qquad ~~~l_{t,u}(s,r;v,Y)= \sum_{m=0}^\infty l_{t,u}^{(m)}(s,r;v) ~Y^m$
    \item Integration over $r$: $\qquad l_{t,u}^{(m)}(s;v) = \int_0^1 \mathrm{d}r ~l_{t,u}^{(m)}(s,r;v)$
    \item Expansion in $v$: $\qquad ~~~\;l_{t,u}^{(m)}(s;v) = \sum_{n=0}^\infty l_{t,u}^{(n,m)}(s;\log v)~ v^n$
    \item Integration over $s$: $\qquad  L_{t,u}^{(n,m)}(\log v) = \int_0^\infty \mathrm{d}s~l_{t,u}^{(n,m)}(s;\log v)$
\end{enumerate}

\paragraph{The integral $\mathbf{L'_0}$.}
The case $L^\prime_0$ brings along various difficulties, most of them associated to the s-channel. The integral to solve is given by
\begin{equation}
\label{Loptu}
   L^\prime_0(x,y,z) = \int_0^1 \mathrm{d}t \int_{0}^{\infty} \mathrm{d}s \int_0^1 \mathrm{d}r ~\frac{ \mathrm{arctanh}~t }{ t \sqrt{(1 + s) (1+t^2 s) }[s r (1-r) x -r y + (1-r) z]},
\end{equation}
or, equivalently, by
\begin{equation}
\label{L0ps}
L^\prime_0(x,y,z) = \int_1^\infty \mathrm{d}\lambda \int_{0}^{\infty} \mathrm{d}s \int_0^1 \mathrm{d}r ~\frac{ \log(1+ \lambda s) }{4 \lambda \sqrt{(1 + s) (1+\lambda s)} [s r (1-r) x -r y + (1-r) z]}.
\end{equation}

\noindent{\it The s-channel.}
The code associated to the s-channel integral $L'_0(v,1-Y,1)$ is not as simple as the ones above, but leads to a comparable performance when implemented in Mathematica. The basic idea is the following: 
The $s$-integral can be split in two parts, an integral from $0$ to $\alpha$ and an integral from $\alpha$ to $\infty$, with $\alpha \gg 1$. Then the integrand to the former integral can be expanded immediately in $v$ and subsequently integrated over, since for $v=0$ the integral (\ref{L0ps}) only diverges at $s\rightarrow \infty$. The integrand of the latter integral is instead expanded for large $s$ (bear in mind that the product $s\times v$ does not have a defined limit), and then integrated over. Eventually, from the sum of the two results $\alpha$ drops out in the limit $\alpha \rightarrow \infty$, yielding the final result.

Let us be more precise and redefine the integrand in eq. (\ref{L0ps}) as
\begin{equation}
    l_s'(s,r,\lambda;v,Y) = a(s,\lambda) b(s,r;v,Y),
\end{equation}
where
\begin{equation}
   a(s,\lambda) = \frac{ \log(1+ \lambda s) }{4 \lambda \sqrt{(1 + s) (1+\lambda s)}}, \qquad b(s,r;v,Y)=\frac{1}{s r (1-r) v -r Y + 1}.
\end{equation}
The procedure is as follows:
\begin{enumerate}
    \item Expansion in $Y$:$\qquad~~~~~~~~~~~\; b(s,r;v,Y) = \sum_{m=0}^\infty b^{(m)}(s,r;v)~Y^m$
     \item Integration region $s \in [\alpha,\infty)$:
    \begin{enumerate}
        \item Expansion in $s$: $\qquad ~~~~\; a(s,\lambda) = \sum_{l=0}^\infty a^{(l)}(\log s,\lambda)~s^{-(l+1)}$
        \item Integration over $s$:$\qquad~\; A^{(m,l)}(\lambda,r,\alpha;v)=\int_\alpha^\infty \mathrm{d}s~ a^{(l)}(\log s,\lambda) b^{(m)}(s,r;v)~s^{-(l+1)}$
        \item Integration over $\lambda$: $\qquad A^{(m,l)}(r,\alpha;v) = \int_1^\infty \mathrm{d}\lambda~A^{(m,l)}(\lambda,r,\alpha;v)$
        \item Expansion in $v$: $\qquad ~~~~A^{(m,l)}(r,\alpha; v) = \sum_{n=0}^\infty A^{(n,m,l)}(r,\alpha;\log v)~v^n $
        \item Integration in $r$: $\qquad~~~ A^{(n,m,l)}(\alpha;\log v) = \int_0^1 \mathrm{d}r~A^{(n,m,l)}(r,\alpha;\log v)$
        \item \label{q1} Summation over $l$: $\qquad A^{(n,m)}(\alpha;\log v)= \sum_{l=0}^n A^{(n,m,l)}(\alpha;\log v)$
    \end{enumerate}
    \item Integration region $s \in [0,\alpha]$ (after the variable substitution $\lambda= t/s$):
    \begin{enumerate}
        \item Expansion in $v$: $\qquad ~~~~~b^{(m)}(s,r;v) = \sum_{n=0}^\infty b^{(n,m)}(s,r)~v^n$
        \item Integration in $r$:$\qquad ~~~~~B^{(n,m)}(s,t)= \int_0^1 \mathrm{d}r~a(s,t)b^{(n,m)}(s,r)$
        \item \label{q2} Integration region $(s,t) \in [0,\alpha]\times [\alpha,\infty)$:
        \vspace{0.15cm}
        \begin{enumerate}
        
        \item Integration in $t$:$\qquad B_1^{(n,m)}(s,\alpha)= \int_\alpha^\infty \mathrm{d}t~B^{(n,m)}(s,t)$
        \item Integration in $s$:$\qquad B_1^{(n,m)}(\alpha)= \int_0^\alpha \mathrm{d}s~B_1^{(n,m)}(s,\alpha)$
         \end{enumerate}
         \vspace{0.15cm}
          \item \label{q3} Integration region $(s,t) \in [0,t]\times [0,\alpha]$:
          \vspace{0.15cm}
          \begin{enumerate}
        \item Integration in $s$:$~~~~~\; B_2^{(n,m)}(t,\alpha)= \int_0^t \mathrm{d}s~B^{(n,m)}(s,t)$
        \item Integration in $t$:$\qquad B_2^{(n,m)}(\alpha)= \int_0^\alpha \mathrm{d}s~B_2^{(n,m)}(s,\alpha)$
         \end{enumerate}
         \vspace{0.15cm}
    \end{enumerate}
    \item Summation over results: $\qquad\, C^{(n,m)}(\alpha;\log v) = A^{(n,m)}(\alpha;\log v)+ B_1^{(n,m)}(\alpha) + B_2^{(n,m)}(\alpha) $
      \item Elimination of $\alpha$: $\qquad~~~~~~~~~ L_s^{\prime(n,m)}(\log v) = \displaystyle\lim_{\alpha \rightarrow \infty}C^{(n,m)}(\alpha;\log v)$
\end{enumerate}
In step (\ref{q1}), the sum clearly runs over all natural numbers. However, for $l>n$, the terms $A^{(n,m,l)}(\alpha;\log v)$ vanish in the limit $\alpha \rightarrow \infty$ and thus can be dropped. Furthermore, note that after the variable substition $\lambda =t/s$ the integration region is $(s,t)\in [0,\alpha]\times [s,\infty)$, which in turn can be divided into two smaller regions. This was done in step (\ref{q2}) and step (\ref{q3}).\\

\noindent{\it The t- and u-channels}.
The implemented code for the t- and u-channels of $L'_0$ is again straightforward. The integrals are respectively given by $L'_0(1-Y,1,v)$ and $L'_0(1,v,1-Y)$, but in this case we start from eq. (\ref{Loptu}).
\begin{enumerate}
    \item Expansion in $Y$: $\qquad ~~~l'_{t,u}(s,r,t;v,Y)= \sum_{m=0}^\infty l_{t,u}^{\prime(m)}(s,r,t;v) ~Y^m$
    \item Integration over $r$: $\qquad l_{t,u}^{\prime(m)}(s,t;v) = \int_0^1 \mathrm{d}r ~l_{t,u}^{\prime(m)}(s,r,t;v)$
    \item \label{probstep} Expansion in $v$: $\qquad ~~~\;l_{t,u}^{\prime(m)}(s,t;v) = \sum_{n=0}^\infty l_{t,u}^{\prime(n,m)}(s,t;\log v)~ v^n$
    \item  Integration over $s$: $\qquad  l_{t,u}^{\prime(n,m)}(t;\log v) = \int_0^\infty \mathrm{d}s~l_{t,u}^{\prime(n,m)}(s,t;\log v)$
    \item Integration over $t$: $\qquad  L_{t,u}^{\prime(n,m)}(\log v) = \int_0^1 \mathrm{d}t~l_{t,u}^{\prime(n,m)}(t;\log v)$
\end{enumerate}
Unfortunately, the Mathematica code does not run efficiently for these channels. However, by constraining oneself in the computation of solely the coefficients linear in $\log v$ in $L_{t,u}^{\prime(n,m)}(\log v)$, the code performance improves drastically. Explicitly, after step (\ref{probstep}), one proceeds as follows:
\begin{enumerate}
    \item[4.] Expansion in $\log v$: $\qquad l_{t,u}^{\prime(n,m)}(s,t;\log v) = l_{t,u}^{\prime(n,m,0)}(s,t)+\log v~ l_{t,u}^{\prime(n,m,1)}(s,t)$ 
    \item[5.]  Integration over $s$: $\qquad  l_{t,u}^{\prime(n,m,1)}(t) = \int_0^\infty \mathrm{d}s~l_{t,u}^{\prime(n,m,1)}(s,t)$
    \item[6.] Integration over $t$: $\qquad  L_{t,u}^{\prime(n,m,1)} =   \int_0^1 \mathrm{d}t~l_{t,u}^{\prime(n,m,1)}(t)$
\end{enumerate}
where $L_{t,u}^{\prime(n,m)}(\log v) = L_{t,u}^{\prime(n,m,0)} + \log v~ L_{t,u}^{\prime(n,m,1)}$. The lack of knowledge of $L_{t,u}^{\prime(n,m,0)}$ is not stringently restrictive for us, since our main interest lies in the anomalous dimensions. Indeed, this only prevents  us from deriving the OPE coefficients of higher weight primaries at second order in the coupling constant.

\section{OPE coefficients}
\label{app:OPE}
\setcounter{equation}{0}
Here we collect some of the OPE coefficients and anomalous dimensions. The (squared) OPE coefficients for the disconnected contribution to the four-point function of a generalized free field of weight $\Delta$ are\footnote{The coefficients can be found in ref. \cite{Fitzpatrick:2011dm}, but we adjusted them to our normalization of conformal blocks.}
\begin{align}{\scriptstyle{
    A_{n,l}=\frac{\pi  \Gamma \left(\frac{d}{2}+l\right) 2^{d-4 \Delta -l-4 n+3} \Gamma \left(-\frac{d}{2}+n+\Delta +1\right) \Gamma (-d+n+2 \Delta +1) \Gamma (l+n+\Delta ) \Gamma (l+2 n+2 \Delta -1) \Gamma \left(-\frac{d}{2}+l+n+2 \Delta \right)}{\Gamma (\Delta )^2 \Gamma (l+1) \Gamma (n+1) \Gamma \left(-\frac{d}{2}+\Delta +1\right)^2 \Gamma \left(\frac{d}{2}+l+n\right) \Gamma \left(-\frac{d}{2}+n+\Delta +\frac{1}{2}\right) \Gamma \left(l+n+\Delta -\frac{1}{2}\right) \Gamma \left(-\frac{d}{2}+l+2 (n+\Delta )\right)}}},\label{genAformula}
\end{align}
where we recall that the conformal blocks are defined via the recursion relations given in ref. \cite{Dolan:2000ut}. They have the general structure
\begin{align*}
    G_{\Delta,l}&= v^{\tfrac{\Delta-l}2} \left(2^{-l} Y^l +Y^{l+1}+...+v(Y^{l-2}+...)+...\right)\mkern-3mu,
\end{align*}
where only the first coefficient, $2^{-l}$, is displayed here, while the others are dropped. This fixes the normalization of the conformal blocks. The recursion begins with the scalar conformal block:
\begin{align}
    G_{\Delta,0}&= v^{\tfrac\Delta2} \tilde{G}(\tfrac{\Delta}2,\tfrac{\Delta}2,\Delta),\\
    \tilde{G}(b,f,S)&=  \sum_{n,m}\frac{Y^m v^n }{m! n! }\frac{ (b)_{m+n} (S-b)_n (f)_{m+n} (S-f)_n}{\left(S-\frac{3}{2}+1\right)_n (S)_{m+2 n}}.
\end{align}

\paragraph{$\mathbf{\Delta=2}$.}
A few OPE coefficients at order $\lambda_R$ are given in table \ref{table1}. Note that, since most of the anomalous dimensions at first order vanish, the OPE coefficients for $l>0$ are determined by matching the OPE expansion at order $\lambda_R^2$. Only the first column, $l=0$, comes from the OPE at order $\lambda_R$.

At order $\lambda_R^2$ the anomalous dimensions of the operators from the first Regge trajectory have a very simple analytic form, see section \ref{chapope}. This seems not to be the case for the subleading trajectories and we simply list some of them in table \ref{table2}. Likewise, few OPE coefficients at order $\lambda_R^2$ can be found (note that due to vanishing of the first order anomalous dimensions of the spinning operators most of the second order OPE coefficients will be only fixed at order $\lambda_R^3$, and we do not have access to) and these are given in table \ref{table3}.

\paragraph{$\mathbf{\Delta=1}$.}
Also for $\Delta=1$ the anomalous dimensions of the operators with nonzero spin vanish at order $\lambda_R$, and thus only corrections to $A_{n,l=0}$ can be determined at this order. These can be found in table \ref{table4}. For $l>0$, the lowest order corrections to the OPE coefficients are determined by matching the OPE expansion at order $\lambda_R^2$. However, owing to the complexity of $L'_4$ at order $\lambda_R^2$, our results are more limited, see table \ref{table5}. 

Anomalous dimensions of the operators at $n>0$ trajectories are purely rational (in terms of $\gamma^2$) and are given in table \ref{table6}. Due to $\gamma^{(1)}_{n,l>0}=0$, again only part of the OPE coefficients can be determined at order $\lambda_R^2$, and the first few of them are given in table \ref{table7}.

\renewcommand{\arraystretch}{1.0} 
\begin{table}[H]
\centering
\begin{tabular}{|c|c|c|c|c|}
\hline
        $\Delta=2$    & $l=0$ & $l=2$ & $l=4$ & $l=6$ \tabularnewline \hline
     $n=0$ & $\frac{1}{3}$ &$\frac{22}{25}$&$\frac{1066}{1323}$&$\frac{1327184}{2760615}$        \tabularnewline
     $n=1$ &  $-\frac{478}{735}$ & $-\frac{20834676}{22204105}$ & $-\frac{4866352}{7966035}$ & $-\frac{13064388928}{43517414655}$        \tabularnewline
     $n=2$ & $-\frac{11507}{50820}$ & $-\frac{14977826}{60144903}$ & $-\frac{15701186862675}{107693096012359}$ & $-\frac{3223599567312}{47662730167195}$\\
     \hline
\end{tabular}
\caption{Some of the OPE coefficients $A^{(1)}_{n,l}$}
\label{table1}
\end{table}
\begin{table}[H]
\centering
\begin{tabular}{|c|c|c|c|c|c|c|}
\hline
       $\Delta=2$     & $l=0$ & $l=2$ & $l=4$ & $l=6$ & $l=8$ & $l=10$ \tabularnewline \hline
     $n=1$ & $\frac{46}{15}$ & $-\frac{107}{1260}$ & $-\frac{19}{1260}$ & $-\frac{131}{27720}$ & $-\frac{301}{154440}$  & $-\frac{19}{20020}$     \tabularnewline
     $n=2$ &  $\frac{113}{28}$ & $-\frac{269}{2520}$ & $-\frac{6707}{311850}$ & $-\frac{1973}{270270}$ & $-\frac{3439}{1081080}$ &        \tabularnewline
     $n=3$ & $\frac{535}{112}$ & $-\frac{6697}{55440}$ & $-\frac{19037}{720720}$ & $-\frac{143581}{15135120}$ & & \\
     \hline
\end{tabular}
\caption{Some anomalous dimensions at order $\lambda_R^2$}
\label{table2}
\end{table}
\begin{table}[H]
\centering
\begin{tabular}{|c|c|c|c|c|c|}
\hline
        $\Delta=2$    & $n=0$ & $n=1$ & $n=2$ & $n=3$ & $n=4$ \tabularnewline \hline
     $l=0$ & $\frac{20}{9}$ & $\frac{111392}{77175}$ & $\frac{27588119}{70436520}$ & $\frac{6664117739}{96718146525}$ & $\frac{54416659121622349}{5688844669692344160}$     \tabularnewline
     \hline
\end{tabular}
\caption{Some of the OPE coefficients $A^{(2)}_{n,l}$}
\label{table3}
\end{table}
\begin{table}[H]
\centering
\begin{tabular}{|c|c|c|c|c|c|}
\hline
        $\Delta=1$    & $n=0$ & $n=1$ & $n=2$ & $n=3$ & $n=4$ \tabularnewline \hline
     $l=0$ & $-2$ & $-\frac{1}{3}$ & $-\frac{617}{36750}$ & $-\frac{20087}{22411620}$ & $-\frac{34519}{695674980}$     \tabularnewline
     \hline
\end{tabular}
\caption{Some of the OPE coefficients $A^{(1)}_{n,0}$}
\label{table4}
\end{table}
\begin{table}[H]
\centering
\begin{tabular}{|c|c|c|}
\hline
        $\Delta=1$    & $l=2$ & $l=4$   \tabularnewline \hline
     $n=0$ & \rule{0pt}{16pt}$\frac{144 \zeta (3)-31}{18 \left(\pi ^2-5\right)}-\frac{43}{15}$ & $\frac{2 (604800 \zeta (3)-380209)}{3675 \left(120 \pi ^2-863\right)}-\frac{10714}{11025}$   \tabularnewline
     \hline
\end{tabular}
\caption{Some of the OPE coefficients $A^{(1)}_{n,l>0}$}
\label{table5}
\end{table}
\begin{table}[H]
\centering
\begin{tabular}{|c|c|c|c|c|c|}
\hline
        $\Delta=1$    & $l=0$ & $l=2$ & $l=4$ & $l=6$ & $l=8$ \tabularnewline \hline
     $n=1$ & $\frac{5}{2}$ & $-\frac{23}{60}$ & $-\frac{59}{420}$ & $-\frac{37}{504}$ & $-\frac{179}{3960}$      \tabularnewline
     $n=2$ &  $\frac{29}{6}$ & $-\frac{373}{1260}$ & $-\frac{71}{630}$ & $-\frac{1693}{27720}$     &   \tabularnewline
     $n=3$ & $\frac{367}{60}$ & $-\frac{641}{2520}$ & $-\frac{15074}{155925}$ & &\\
     \hline
\end{tabular}
\caption{Some anomalous dimensions at order $\lambda_R^2$}
\label{table6}
\end{table}
\begin{table}[H]
\centering
\begin{tabular}{|c|c|c|c|}
\hline
       $\Delta=1$     & $n=0$ & $n=1$ & $n=2$  \tabularnewline \hline
     $l=0$ & $-4 \zeta (3)+\pi ^2/2+8$ & $\frac{37}{27}$ & $\frac{337219}{7717500}$  \tabularnewline
     \hline
\end{tabular}
\caption{Some of the OPE coefficients $A^{(2)}_{n,l}$}
\label{table7}
\end{table}

\end{appendix}

\newpage
\providecommand{\href}[2]{#2}\begingroup\raggedright\endgroup

\end{document}